\documentclass[aps,prd,twocolumn,nofootinbib,superscriptaddress,preprintnumbers,floatfix]{revtex4-1}
\usepackage{lipsum}
\usepackage{graphicx}
\usepackage{color}
\usepackage{enumerate}
\usepackage{amsmath}
\usepackage{bm}
\usepackage{bbold}
\usepackage{comment}
\usepackage{diagbox}
\usepackage{amssymb}
\usepackage{stmaryrd}
\usepackage{multirow}
\usepackage[normalem]{ulem}
\usepackage{float}
\usepackage{longtable,booktabs}
\usepackage[dvipsnames,table,xcdraw]{xcolor}
\usepackage[normalem]{ulem}
\usepackage{physics}
\usepackage{url}
\usepackage{soul} 
\usepackage{verbatim}
\usepackage{siunitx}
\usepackage{booktabs,dcolumn}
\usepackage{hyperref}
\usepackage{cleveref}
\usepackage{pythonhighlight}

\Crefname{equation}{Eq.}{Eqs.}
\Crefname{section}{Section}{Sections}
\Crefname{figure}{Fig.}{Figs.}
\Crefname{appendix}{Appendix}{Appendices}

\newcommand{\pislash}[0]{\pi\!\!\!/}

\definecolor{darkblue}{rgb}{0.1,0.2,0.6} \definecolor{darkred}{rgb}{0.8,0.1,0.2}

\usepackage{multirow}
\usepackage[caption=false]{subfig}

\def\beq{\begin{equation}}
\def\eeq{\end{equation}}

\begin{document}
\long\def\/*#1*/{}

\title{ Constraint of pionless EFT using two-nucleon spectra from lattice QCD }

\author{William~Detmold} 
\affiliation{
	Center for Theoretical Physics, 
	Massachusetts Institute of Technology, 
	Cambridge, MA 02139, USA}
\affiliation{The NSF AI Institute for Artificial Intelligence and Fundamental Interactions}

\author{Fernando Romero-L\'opez} 
\affiliation{
	Center for Theoretical Physics, 
	Massachusetts Institute of Technology, 
	Cambridge, MA 02139, USA}
\affiliation{The NSF AI Institute for Artificial Intelligence and Fundamental Interactions}

\author{Phiala~E.~Shanahan} 
\affiliation{
	Center for Theoretical Physics, 
	Massachusetts Institute of Technology, 
	Cambridge, MA 02139, USA}
\affiliation{The NSF AI Institute for Artificial Intelligence and Fundamental Interactions}

\preprint{MIT-CTP/5538}

\begin{abstract}
Finite-volume pionless effective field theory (FVEFT$_{\pislash}$) at next-to-leading order (NLO) is used to analyze the two-nucleon lattice QCD spectrum of Ref.~\cite{Amarasinghe:2021lqa}, performed at quark masses corresponding to a pion mass of approximately $800 $ MeV. Specifically, the effective theory is formulated in finite volume, and variational sets of wave functions are optimized using differential programming. Using these wave functions projected to the appropriate finite-volume symmetry group, variational bounds from FVEFT$_{\pislash}$ are obtained for the ground state, as well as excited states. By comparison with the lattice QCD GEVP spectrum, different low energy constants (LECs) are constrained. Relativistic corrections are incorporated, allowing for the extractions of NLO LECs, as well as the leading $s$-$d$-wave mixing term in the deuteron channel.
\end{abstract}
 
\maketitle

\section{Introduction}

Nuclear physics originates from the nonperturbative dynamics of Quantum Chromodynamics (QCD), the theory of strong interaction between quarks and gluons. However, first-principles predictions of nuclear properties remain a complicated task. Lattice QCD (LQCD) offers a systematically improvable approach to perform calculations in QCD. In particular, recent progress in the field has enabled computations in few-nucleon systems~\cite{Beane:2003da,Beane:2006mx,Beane:2009gs,Doi:2011gq,NPLQCD:2012mex,Yamazaki:2012hi,Orginos:2015aya,Berkowitz:2015eaa,Yamazaki:2015asa,Wagman:2017tmp,Francis:2018qch,Horz:2020zvv,NPLQCD:2020lxg,Green:2021qol,Amarasinghe:2021lqa}. 
Nonetheless, achieving controlled calculations of many-nucleon systems in LQCD remains an open problem, given the difficulties to overcome the signal-to-noise problem and the growing computational cost of Wick contractions in these systems.

Complementary to direct LQCD calculations of the properties of  larger nuclei, matching of LQCD to effective field theories (EFTs) and nuclear many-body methods provides another approach to describing the properties of nuclei using first-principles QCD information. This can be realized in several ways. One option is to match infinite-volume extrapolated LQCD data to an EFT, and then use LQCD instead of experimental constraints in EFT and many-body pipelines~\cite{Kamada:2001tv,Barnea:2013uqa,Kirscher:2015yda,Carlson:2014vla,Bansal:2017pwn,Gandolfi:2017arm,Contessi:2017rww,Epelbaum:2010xt,Lu:2019nbg,Lahde:2019npb}. A promising alternative is to directly match finite-volume LQCD calculations to an EFT formulated in the same finite volume, and then perform many-body computations within the same EFT directly in infinite volume~\cite{Detmold:2004qn,Briceno:2012yi,Eliyahu:2019nkz,Detmold:2021oro,Sun:2022frr}. An important advantage of the latter approach is that it avoids the need for an infinite-volume extrapolation in LQCD calculations, and makes maximal use of LQCD results at fixed volume.

Pionless EFT (EFT$_{\pislash}$) is an effective theory for nucleons that is valid at low momenta~\cite{Kaplan:1996xu,Kaplan:1998tg,Kaplan:1998we,vanKolck:1998bw,Bedaque:1998kg,Chen:1999tn,Bedaque:2002yg,Bedaque:2002mn} (see Ref.~\cite{Hammer:2019poc} for a recent review). The expansion parameter of EFT$_{\pislash}$ is $k^2/M_\pi^2$, where $M_\pi$ is the pion mass and $k$ the nucleon momentum, and its radius of convergence is set by the $t$-channel cut, located at $|k^2|=M_\pi^2/4$.
EFT$_{\pislash}$ can be expressed as a tower of contact interactions between nucleon fields each accompanied by a corresponding Low Energy Constant (LEC), and it can be formulated both in finite and infinite volume~\cite{Barnea:2013uqa,Eliyahu:2019nkz,Detmold:2021oro,Bazak:2022mjh}. With LECs fixed by LQCD or experiment, calculations in EFT$_{\pislash}$ have been performed for systems as large as atomic number $A=40$~\cite{Bansal:2017pwn}.

In recent work, finite-volume EFT$_{\pislash}$ (FVEFT$_{\pislash}$) at leading order has been used~\cite{Eliyahu:2019nkz,Detmold:2021oro,Sun:2022frr} to  analyze the LQCD calculations of Ref.~\cite{NPLQCD:2012mex} (see also Refs.~\cite{Beane:2003da,Beane:2003yx,Beane:2012ey,Li:2019qvh,Li:2021mob,Severt:2022jtg} for other works on EFTs in finite volume).
By optimizing variational wave-function ans\"atze, the LECs of the theory were obtained after matching finite-volume LQCD and EFT$_{\pislash}$ energies and matrix elements. With that set of LECs, infinite-volume predictions for systems of up to six nucleons were evaluated. 
It was subsequently shown in Ref.~\cite{Sun:2022frr} that differential programming and machine learning provide an efficient way of optimizing variational wave functions. 

This work further extends the application of FVEFT$_{\pislash}$, and shows that
it can be used to analyze a finite-volume LQCD spectrum including excited states as well as states in different irreducible representations of the finite-volume symmetry group and with nonzero values of the total momentum of the system. As such, it is shown how FVEFT$_{\pislash}$ can be used as an alternative to finite-volume multi-particle formalisms based on quantization conditions~\cite{Luscher:1986pf,Hansen:2014eka,Hansen:2015zga,Hammer:2017uqm,Hammer:2017kms,Mai:2017bge,Draper:2023xvu,Bubna:2023oxo}.
Additional terms are implemented in the EFT Hamiltonian with respect to previous work~\cite{Detmold:2021oro,Sun:2022frr}, specifically operators with up to two derivatives and the first relativistic corrections. 
Following the approach of Ref.~\cite{Sun:2022frr}, several sets of correlated Gaussian wave functions are optimized. By matching to the variational LQCD energy bounds from Ref.~\cite{Amarasinghe:2021lqa}, the leading order (LO) and next-to-leading order (NLO) LECs are obtained, and the effects of both the LO and NLO couplings are found to be statistically significant. The magnitude of the $s$-$d$-wave mixing term in the deuteron channel is also bounded by consideration of differences between energy levels in different irreducible representations of the cubic group.

This paper is organized as follows. \Cref{sec:FVEFT} introduces the necessary background to formulate pionless EFT in a finite volume: \Cref{subsec:pionlessEFTNLO} describes the EFT at NLO, \Cref{subsec:HamEFT} introduces the Hamiltonian formulation and the regularization of the interactions, \Cref{subsec:variational} discusses the variational approach to FVEFT$_{\pislash}$, and \Cref{subsec:numerical} provides a toy numerical demonstration of the effect of the different terms of the Hamiltonian in the variational spectrum. The application of FVEFT$_{\pislash}$ to constrain the LECs using LQCD data from Ref.~\cite{Amarasinghe:2021lqa} is discussed in \Cref{sec:appLQCD}, at LO in \Cref{subsec:LO}, at NLO in \Cref{subsec:NLO}, and including $s$-$d$-wave mixing in \Cref{subsec:sd}. \Cref{sec:conclusion} presents a conclusion and summary. \Cref{app:matrix} collects the necessary expressions to evaluate the matrix elements of the Hamiltonian used in this work.

\section{Pionless EFT in a finite volume}
\label{sec:FVEFT}
Pionless EFT~\cite{Kaplan:1996xu,Kaplan:1998tg,Kaplan:1998we,vanKolck:1998bw,Bedaque:1998kg,Chen:1999tn,Bedaque:2002yg,Bedaque:2002mn} describes the interactions between nucleons in an energy range in which no other degrees of freedom are dynamical, i.e., pions, and other meson and baryon resonances have been integrated out. The radius of convergence of the EFT is limited by the $t$-channel cut, which originates from processes where two nucleons exchange a pion. Therefore, the relative momentum in the two-nucleon system is restricted to be in the region $|k^2| < M_\pi^2/4$ for convergence.

In this section, the procedure to constrain LECs by performing a direct matching of the energy levels between LQCD and pionless EFT at NLO in a finite volume is discussed.

\subsection{Pionless EFT at NLO}
\label{subsec:pionlessEFTNLO}

The EFT$_{\pislash}$ Lagrangian including terms up to quadratic order in momentum, and including at most two-nucleon interactions, is given by~\cite{Rupak:1999rk}:
\begin{equation}
    \mathcal L_{\pislash} = \mathcal{L}_K + \mathcal{L}_{2}^\text{LO} + \mathcal{L}_{2}^\text{NLO} +  \mathcal{L}_{2}^{sd}   + \hdots.
\end{equation}
This Lagrangian describes both the deuteron, with spin $S=1$ and isospin $I=0$, and the dineutron, with $S=0$ and $I=1$. Here, $\mathcal L_K$ refers to the kinetic term, and $ \mathcal{L}_{2}^\text{LO}$ and $ \mathcal{L}_{2}^\text{NLO}$ to the leading and next-to-leading $s$-wave interactions.  In addition, the $s$-$d$-wave mixing term for the deuteron channel, $\mathcal{L}_{2}^{sd}$, is also included, since it also contains two derivatives and will affect the finite-volume energies that are used in this work.

In the kinetic term, the first relativistic correction is considered, given that the aim is to analyze excited states in the two-nucleon spectrum:
\begin{equation}
     \mathcal{L}_K  = N^{\dagger}\left(i D_0+\frac{\mathbf{D}^2}{2 M_N}  + \frac{\mathbf{D}^4}{8M_N^3} +  \hdots \right) N , 
\end{equation}
where $N$ is the nucleon field, $M_N$ is the nucleon mass, $D_0$ is the temporal derivative and $\mathbf D$ is the vector of spatial derivatives. 

At LO, the interaction Lagrangian is given by
\begin{align}
    \begin{split}
     \mathcal{L}_{2}^\text{LO}  =&   -C_S\left(N^T P_i N\right)^{\dagger}\left(N^T P_i N\right) \\
& -C_T\left(N^T \bar{P}_a N\right)^{\dagger}\left(N^T \bar{P}_a N\right),
    \end{split}
\end{align}
where $C_S$ and $C_T$ are the LECs corresponding to the deuteron and dineutron channel. Repeated indices denote summation. The deuteron and dineutron projectors are defined as
\begin{equation}
P_i \equiv \frac{1}{\sqrt{8}} \sigma_2 \sigma_i \tau_2, \quad \bar{P}_a \equiv \frac{1}{\sqrt{8}} \sigma_2 \tau_2 \tau_a,
\label{eq:proj}
\end{equation}
where $\sigma_i$ ($\tau_a$) are Pauli matrices in spin (isospin) space.

At next-to-leading order (NLO), the following terms contribute
\begin{align}
    \begin{split}
      \mathcal{L}_{2}^\text{NLO} =&   -C^{(2)}_S\left(N^T \mathcal{O}^{(2)}_i N\right)^{\dagger}\left(N^T P_i N\right) + \text{ h.c. } \\
       &-C^{(2)}_T\left(N^T  \bar{\mathcal{O}}^{(2)}_a N\right)^{\dagger}\left(N^T \bar{P}_a N\right) + \text{ h.c. },
    \end{split}
\end{align}
where h.c. denotes the Hermitian conjugate, $C^{(2)}_S$ and $C^{(2)}_T$ are LECs, the operator
\begin{equation}
\mathcal{O}_i^{(2)}=\frac{1}{4}\left[\overleftarrow{\mathbf{D}}^2 P_i-2 \overleftarrow{\mathbf{D}} P_i \overrightarrow{\mathbf{D}}+P_i \overrightarrow{\mathbf{D}}^2\right],
\label{eq:opO2}
\end{equation}
and $\bar{\mathcal{O}}_a^{(2)}$ is defined analogously for the dineutron channel with the replacement $P_i \to \bar P_a$. In \Cref{eq:opO2}, the arrows over the derivatives indicate whether they act on the fields to the left or to the right.

Moreover, in the deuteron channel at NLO one must include a term that mixes $s$ and $d$ waves:
\begin{equation}
     \mathcal{L}_{2}^{sd} = 
C^{(s d)} \left(N^T P_i N\right)^{\dagger}\left(N^T \mathcal{O}^{kl}_{j} N\right) \mathcal{T}^{i j}_{kl}+ \text{ h.c. } , 
\end{equation}
where $C_2^{(s d)}$ is the LEC that controls this mixing, 
\begin{equation}
\mathcal{O}^{kl}_{j}=\overleftarrow{\mathbf{D}}^k \overleftarrow{\mathbf{D}}^l P_j-\overleftarrow{\mathbf{D}}^k P_j \overrightarrow{\mathbf{D}}^l-\overleftarrow{\mathbf{D}}^l P_j \overrightarrow{\mathbf{D}}^k+P_j \overrightarrow{\mathbf{D}}^k\overrightarrow{\mathbf{D}}^l,
\end{equation}
and 
\begin{equation}
\mathcal{T}_{k l}^{ij}=\left(\delta^{i k} \delta^{j l}-\frac{1}{3} \delta^{i j} \delta^{k l}\right).
\end{equation}

Finally, interactions of three or more nucleons can also be included, but they will not be used in the present work, which focuses on two-nucleon systems. 

\subsection{Hamiltonian for the EFT}
\label{subsec:HamEFT}

In this section, the Hamiltonian formulation of the EFT presented in \Cref{subsec:pionlessEFTNLO} is discussed. The Hamiltonian is constructed as
\begin{align}
\begin{split}
H&=\sum_n K_n  + \sum_{n<m} V_2\left(\mathbf{r}_{n m}\right), 
\label{eq:hamKV}
\end{split}
\end{align}
where $n,m \in \{1,..,A\}$ label the nucleons in an $A$-nucleon system. Here, only systems of two nucleons are considered, and the sum $n<m$ can only take one value. The sum is thus omitted henceforth to simplify notation. Furthermore, note that in two-nucleon systems, each two-body isospin channel can be considered independently. 

The kinetic energy operator is 
\begin{equation}
  K_n = K_n^{(2)} +  K_n^{(4)} = \left[ -  \frac{1}{2 M_N}  \overrightarrow{\boldsymbol \nabla}_n^2 - \frac{1}{8 M^3_N}  ( \overrightarrow{\boldsymbol \nabla}_n^2)^2 \right], \label{eq:kineticH}
\end{equation}
where $  \overrightarrow{\boldsymbol \nabla}_n$ is the vector derivative with respect to the coordinates of the $n$-th particle, and $ \overrightarrow{\boldsymbol\nabla}_n^2$ is the corresponding Laplacian. In \Cref{eq:kineticH}, $K_n^{(2)}$ and  $K_n^{(4)}$ are implicitly defined to be the nonrelativistic kinetic energy, and the first relativistic correction, respectively. 

The two-particle interaction potential, $V_2\left(\mathbf{r}_{n m}\right)$, is a function of the displacements between particles $n$ and $m$, ${\mathbf r_{nm} =\mathbf{r}_n - \mathbf{r}_m}$, where $\mathbf{r}_n=(r_n^{(x)}, r_n^{(y)}, r_n^{(z)})$ labels the position of the $n$-th particle. The two-nucleon interactions are regulated with Gaussian smearing~\cite{PhysRevA.87.063609}. In infinite volume, the regulator is
\begin{equation}
\begin{aligned}
g_{\Lambda}(\mathbf{r}) & =\frac{\Lambda^3}{8 \pi^{3 / 2}} \exp \left(-\Lambda^2|\mathbf{r}|^2 / 4\right) \\
& =\frac{\Lambda^3}{8 \pi^{3 / 2}} \prod_{\alpha \in\{x, y, z\}} \exp \left(-\frac{\Lambda^2}{4} r^{(\alpha) 2} \right) ,
\end{aligned}
\end{equation}
where $\Lambda$ is a the regulator parameter, which can be related to a length scale $r_0$ as $\Lambda = \sqrt{2}/r_0$. This choice constitutes a family of renormalization schemes for the LECs in the Hamiltonian, and the concrete choice of $\Lambda$ (or equivalently $r_0$) specifies the scheme.

At leading order, 
\begin{align}
    \begin{split}
    V^\text{LO}_{2} \left(\mathbf{r}_{nm}\right)=& \,C_S     P'_i \,   P_i   \,   g_{\Lambda}\left(\mathbf{r}_{nm}\right) \\
    &+ C_T  \bar P'_a  \,  \bar P_a    \, g_{\Lambda}\left(\mathbf{r}_{nm}\right), \label{eq:V2LOginf}
    \end{split}
\end{align}
where the projectors $P$ and $\bar P$ act on the incoming state, and the $P'$ and $\bar P'$ on the outgoing state. The $s$-wave NLO potential is:
\begin{align}
    \begin{split}
    V^{\text{NLO}}_{2} \left(\mathbf{r}_{nm}\right)=& \left( C^{(2)}_S  P'_i   P_i + C^{(2)}_T  \bar P'_a \,  \bar P_a \right)  \\ & \times \Bigg[  g_{\Lambda}\left(\mathbf{r}_{nm}  \right)  \big(   \overrightarrow{\boldsymbol \nabla}_n - \overrightarrow{\boldsymbol \nabla}_m  \big)^2  
     \\ &+ \big(   \overleftarrow{\boldsymbol \nabla}_n - \overleftarrow{\boldsymbol \nabla}_m  \big)^2  g_{\Lambda}\left(\mathbf{r}_{nm}  \right)
    \Bigg].
    \label{eq:V2NLOpot}
    \end{split}
\end{align}
 Here, the right or left arrows above the $\boldsymbol \nabla$ operator indicate whether it acts on the initial or final states, respectively. Note that in the scheme used here, the derivative operators never act on the Gaussian regulator function. Finally, for the deuteron, the $s$-$d$-wave mixing term is specified as
\begin{align}
    \begin{split}
    V^{sd}_{2} \left(\mathbf{r}_{nm}\right)&= C^{(sd)}\mathcal{T}_{k l}^{ij}   P'_i  P_j \times \\ &\Bigg[  g_{\Lambda}\left(\mathbf{r}_{nm}  \right)  \big(   \overrightarrow{\nabla}^{(k)}_n - \overrightarrow{\nabla}^{(l)}_m  \big)^2   \\
    & + \big(   \overleftarrow{\nabla}^{(k)}_n - \overleftarrow{\nabla}^{(l)}_m  \big)^2  g_{\Lambda}\left(\mathbf{r}_{nm}  \right)
    \Bigg],
    \label{eq:sdpotential}
    \end{split}
\end{align}
where ${\nabla}^{(k)}_n$ is the $k$-th component of $\boldsymbol \nabla_n$, i.e., the derivative with respect to $r_n^{(k)}$.

Finally, the Hamiltonian can be formulated in a finite periodic box of side-length $L$. In practice, the only necessary change involves the regulator, for which periodicity can be imposed by  summing over copies translated by multiples of $L$ in each spatial direction,
\begin{equation}
\begin{aligned}
g_{\Lambda}(\mathbf{r}, L)&=  \frac{\Lambda^3}{8 \pi^{3 / 2}} \times  \\
 \prod_{\alpha \in\{x, y, z\}} & \sum_{q^{(\alpha)}=-q_\text{cut}}^{q_\text{cut}} \exp \left(-\frac{\Lambda^2}{4}\left(r^{(\alpha)}-L q^{(\alpha)}\right)^2\right).
\label{eq:glambdaFV}
\end{aligned}
\end{equation}
Exact periodicity is achieved as $q_\text{cut} \to \infty$. In practice, the magnitude of contributions decays exponentially with $q^{(\alpha)}$, and in numerical calculations a finite value of $q_\text{cut}$ is sufficient to compute results to any given numerical precision.
The finite-volume potential is then given by \Cref{eq:V2LOginf,eq:V2NLOpot,eq:sdpotential} under the replacement ${g_{\Lambda}(\mathbf{r}) \to g_{\Lambda}(\mathbf{r}, L)}$.

\subsection{Variational approach}
\label{subsec:variational}

The variational method is a systematically improvable approach to determining increasingly restrictive upper bounds on some set of energies in a quantum system. For the ground state, it proceeds as follows. Given a wave function ansatz $\Psi$ with some definite quantum numbers, the expectation value of the energy for that wave function, $\mathcal{E}\left[\Psi\right]$, constitutes an upper bound of the ground state energy $E_0$ in that sector of quantum numbers:
\begin{equation}
E_0 \leq \mathcal{E}\left[\Psi\right]=\frac{ \left \langle \Psi | H | \Psi \right \rangle }{\left \langle \Psi | \Psi \right \rangle}.
\end{equation}
Note that for simplicity, internal quantum numbers such as spin or isospin are kept implicit in the wave function.

This can be generalized to energy levels beyond the ground state; given a set of `trial' wave functions $\{\Psi_i \}$ with ${i \in \{1,..., N_g\}}$, and solving the generalized eigenvalue problem (GEVP):
\begin{equation}
\mathbb{H}\, \mathbf{v}= \lambda\, \mathbb{N} \,\mathbf{v},
\label{eq:GEVP}
\end{equation}
where 
\begin{equation}
\left[\mathbb{N}\right]_{i j} \equiv \left \langle \Psi_i | \Psi_j \right \rangle, \quad \text{and} \quad \left[\mathbb{H}\right]_{i j} \equiv \left \langle \Psi_i | H | \Psi_j \right \rangle,
\end{equation}
yields $N_g$ eigenvalues $\lambda_1, ..., \lambda_{N_g}$, ordered by increasing energy, which correspond to upper bounds of the lowest $N_g$ energy levels. 
Moreover, the eigenvector corresponding to the $i$-th eigenvalue is $\mathbf{c}_{i} = (c^{(1)}_{i}, ..., c^{(N_g)}_{i} )$, which can be used to construct an approximate wave function for the corresponding state,  $\sum_j  c^{(j)}_i \ket{\Psi_j}$. This approximation, i.e. a `variational' wave function, should approach the true wave function as the set of trial wave functions approaches a basis for the relevant Hilbert space.

\subsubsection{Variational wave functions }

In this section, the trial functions used in this work are described. Following previous work~\cite{Detmold:2021oro,Sun:2022frr}, a factorization of the spin-isospin and the spatial parts is assumed:
\begin{equation}
    \ket{\Psi} = \Psi(\mathbf x) \ket{\chi_h} .
\end{equation}
Here $\mathbf x$ (with components $x_n^{(\alpha)}$) describes the positions of all particles in the three spatial dimensions, and $h$ is a generic label for spin-isospin quantum numbers.

The spin-isospin states for the two-nucleon systems are chosen to be antisymmetric and given by:
\begin{align}
    \begin{split}
\left|\chi_{d, S_z=+1}\right\rangle & =\frac{1}{\sqrt{2}}\left[\left|p^{\uparrow} n^{\uparrow}\right\rangle-\left|n^{\uparrow} p^{\uparrow}\right\rangle\right], \\
\left|\chi_{d, S_z=0}\right\rangle & =\frac{1}{2}\left[\left|p^{\uparrow} n^{\downarrow}\right\rangle-\left|n^{\uparrow} p^{\downarrow}\right\rangle+\left|p^{\downarrow} n^{\uparrow}\right\rangle-\left|n^{\downarrow} p^{\uparrow}\right\rangle\right], \\
\left|\chi_{n n}\right\rangle & =\frac{1}{\sqrt{2}}\left[\left|n^{\uparrow} n^{\downarrow}\right\rangle-\left|n^{\downarrow} n^{\uparrow}\right\rangle\right]. \label{eq:spiniso}
    \end{split}
\end{align}
 Here $n,p$ denote neutron and proton one-particle states, and the superscript arrows correspond to the third component of the spin. Moreover, $\ket{\chi_{nn}}$ indicates the dineutron channel, and $\ket{\chi_{d,S_z}}$ the deuteron channel with third component of the two-nucleon spin $S_z$. Other spin-isospin states not displayed here, e.g. $\left|\chi_{pp}\right\rangle$, can be defined analogously.

The spatial wave functions are assumed to factorize:
\begin{equation}
    \Psi(\mathbf x) =  \prod_{\alpha \in \{x,y,z\}}  \Psi^{(\alpha)}(\mathbf x^{(\alpha)}).
\end{equation}
Then, correlated Gaussian ans\"atze are considered independently for each component. In infinite volume, these are
\begin{equation}
\begin{gathered}
\Psi_{\infty}^{(\alpha)}\left(A^{(\alpha)}, B^{(\alpha)}, \mathbf{d}^{(\alpha)} ; \mathbf{x}^{(\alpha)}\right)=\exp \left[-\frac{1}{2} \mathbf{x}^{(\alpha) T} A^{(\alpha)} \mathbf{x}^{(\alpha)}\right. \\
\left.-\frac{1}{2}\left(\mathbf{x}^{(\alpha)}-\mathbf{d}^{(\alpha)}\right)^T B^{(\alpha)}\left(\mathbf{x}^{(\alpha)}-\mathbf{d}^{(\alpha)}\right)\right],
\end{gathered}
\end{equation}
where $A^{(\alpha)},B^{(\alpha)}$, and  $\mathbf{d}^{(\alpha)}$ contain the free parameters of the ansatz. Specifically,  $A^{(\alpha)}$ and $B^{(\alpha)}$ are $N_n \times N_n$ real symmetric matrices where $N_n$ is the number of particles, and $\mathbf{d}^{(\alpha)}$ is an $N_n$-component real-valued vector. In finite volume, periodic boundary conditions are imposed by summing periodically-translated copies of the wave function ansatz:
\begin{equation}
\begin{aligned}
& \Psi_L^{(\alpha)}\left(A^{(\alpha)}, B^{(\alpha)}, \mathbf{d}^{(\alpha)} ; \mathbf{x}^{(\alpha)}\right)= \\
& \sum^{\mathbf{b}_\text{cut}}_{\mathbf{b}^{(\alpha)}=-\mathbf{b}_\text{cut}} \Psi_{\infty}^{(\alpha)}\left(A^{(\alpha)}, B^{(\alpha)}, \mathbf{d}^{(\alpha)} ; \mathbf{x}^{(\alpha)}-\mathbf{b}^{(\alpha)} L\right),
\label{eq:psiLalpha}
\end{aligned}
\end{equation}
where $\mathbf{b}^{(\alpha)}$ is an integer-valued vector with $N_n$ components for each $\alpha$. Similar to \Cref{eq:glambdaFV}, exact periodicity requires an infinite sum, but in practice each component of $\mathbf{b}_\text{cut}$ need only be large enough that \Cref{eq:psiLalpha} converges to the desired numerical precision. 
Finally, to respect Fermi statistics, the spatial wave function is forced to be symmetric by summing over all possible permutations $\mathcal P$ of the particles:
\begin{equation}
\Psi_L^{\mathrm{sym}}(A, B, \mathbf{d} ; \mathbf{x})=\sum_{\mathcal{P}} \Psi_L\left(A_{\mathcal{P}}, B_{\mathcal{P}}, \mathbf{d}_{\mathcal{P}} ; \mathbf{x}\right),
\end{equation}
where $A= (A^{(x)}, A^{(y)}, A^{(z)} )$, and similarly for $B$ and $\mathbf d $, and the permutation acts by swapping rows and columns of the matrices and entries of the vectors. For example,
\begin{align}
\begin{split}
    \left(A^{(\alpha)}_{\mathcal P}\right)_{mn} &=  \left(A^{(\alpha)}\right)_{\mathcal P(m) \mathcal P(n)} \, , \\ \left(d^{(\alpha)}_{\mathcal P}\right)_{n} &= \left(d^{(\alpha)}\right)_{\mathcal P(n)} \, ,
    \end{split}
\end{align}
where $\mathcal P(n)$ is the $n$-th element of the permutation $\mathcal P$.

In summary, the finite-volume wave function ans\"atze are a linear combination of $N_g$ wave functions:
\begin{equation}
     \ket{\Psi^{(h)}}  = \sum_{i=1}^{N_g} c_i  \ket{\Psi^{(h)}_i}, 
     \label{eq:lincombansatz}
\end{equation}
where the $i$-th term looks like
\begin{equation}
  \ket{\Psi^{(h)}_i} =  \Psi_L^{\mathrm{sym}}(A_i, B_i, \mathbf{d}_i ; \mathbf{x}) \ket{\chi_h} \equiv  \Psi_{L,i}^{\mathrm{sym}}(\mathbf x)\ket{\chi_h},
  \label{eq:ansatz}
\end{equation}
for each choice of spin-isospin quantum numbers, $h$.

\subsubsection{Evaluation of matrix elements}
\label{subsec:matels}

In order to solve the GEVP in \Cref{eq:GEVP} given some set of trial wave functions
and thereby obtain variational bounds on the spectrum, the matrix elements of both $\mathbb{N}$ and $\mathbb{H}$ need to be evaluated.
A convenient consequence of the choice of a correlated Gaussians ansatz is that all of the required matrix elements can be evaluated analytically, as they all reduce to multi-dimensional Gaussian integrals.\footnote{See Table 7.1 in Ref.~\cite{Suzuki:1998bn} for useful expressions in the context of multi-dimensional Gaussian integrals.} This section provides an overview of the necessary pieces, while the explicit expressions are collected in \Cref{app:matrix}.

In order to obtain variational bounds, the normalization is needed
\begin{equation}
 \left[\mathbb{N}\right]_{i j} =  \delta_{h h'} \int d \mathbf x \, \Psi_{L,i}^{\mathrm{sym}}(\mathbf x)  \Psi_{L,j}^{\mathrm{sym}}(\mathbf x).
 \label{eq:Nsymm}
\end{equation}
For notational simplicity, the $h$ indices are omitted on the  left-hand side of this and the following equations.
For the Hamiltonian, several terms need to be evaluated:
\begin{equation}
    \left[\mathbb{H}\right]_{i j} = \left[\mathbb{K}^{(2)} +\mathbb{K}^{(4)}+\mathbb{V}^\text{LO}_2 + \mathbb{V}^\text{NLO}_2 + \mathbb{V}^{sd}_2\right]_{i j}.
    \label{eq:Hij}
\end{equation}
Here $\mathbb{K}^{(2)}$ represents matrix elements of the nonrelativistic kinetic energy:
\begin{equation}
 \left[\mathbb{K}^{(2)}\right]_{i j} =  \delta_{h h'} \sum_n \int d \mathbf x \, \Psi_{L,i}^{\mathrm{sym}}(\mathbf x)  K^{(2)}_n \Psi_{L,j}^{\mathrm{sym}}(\mathbf x),
\end{equation}
with $K^{(2)}_n$ given in \Cref{eq:kineticH}, and similarly for the first relativistic correction, $\mathbb{K}^{(4)}$. Next, 
$\mathbb{V}_2^\text{LO}$ is constructed from the matrix elements of the LO potential:
\begin{equation}
 \left[\mathbb{V}^\text{LO}_2\right]_{i j} =  \delta_{h h'} \int_{ d \mathbf x} \, \Psi_{L,i}^{\mathrm{sym}}(\mathbf x)  \langle \chi_h| V_2^\text{LO}| \chi_h \rangle \Psi_{L,j}^{\mathrm{sym}}(\mathbf x),
\end{equation}
with a similar expression for the NLO potential, $\mathbb{V}^\text{NLO}_2$. These terms do not mix spin or isospin components, which is made explicit by $\delta_{h h'}$, indicating that the matrix elements vanish unless all isospin/spin quantum numbers are identical in the initial and final states.
By contrast, the expression for the $s$-$d$-wave mixing term mixes the spin components of the deuteron:
\begin{equation}
    \left[\mathbb{V}^{sd}_2\right]_{i j} = \int_{ d \mathbf x} \, \Psi_{L,i}^{\mathrm{sym}}(\mathbf x)  \langle \chi_{h'}| V_2^\text{LO}| \chi_{h} \rangle \Psi_{L,j}^{\mathrm{sym}}(\mathbf x).
\end{equation}
By choosing a Cartesian basis for the spin components,  $\langle \chi_{h'}| V_2^{sd}| \chi_{h} \rangle $ is easily related to the $T_{kl}^{ij}$ tensor in \Cref{eq:sdpotential}.

After solving the GEVP in \Cref{eq:GEVP} to determine energy bounds and approximate eigenvectors, one can evaluate observables using variational wave functions, e.g.  matrix elements as in Ref.~\cite{Detmold:2021oro}. One observable that is particularly useful in the classification of states is the the total three-momentum $\mathbf P$ of the multi-particle system.  In a finite volume, $\mathbf P$ is quantized as $(2\pi/L) \mathbf n $ with $\mathbf{n} \in \mathbb Z^3$, and each value defines a ``frame'' such as, the rest frame with $\mathbf P^2=0$. In LQCD spectroscopy, it is customary to compute energy levels at several different definite values of $\mathbf{P}^2$. For this reason, it will be useful to evaluate the expectation value of the $\mathbf P^2$ operator on the variational wave functions. For this, the following expression is needed
\begin{equation}
      \left[\mathbb{P}^2\right]_{i j} = \delta_{h h'} \int d \mathbf x \, \Psi_{L,i}^{\mathrm{sym}}(\mathbf x) \mathbf P^2 \Psi_{L,j}^{\mathrm{sym}}(\mathbf x),
\end{equation}
where $\mathbf P^2 = \left(\sum_n \mathbf P_n\right)^2$, and $\mathbf P_n = -i \boldsymbol \nabla_n$.

\subsubsection{Optimization of the wave functions}

Efficient use of the variational approach relies on having wave function ans\"atze that have the flexibility to provide accurate representations of the eigenstates. These can be difficult to construct, and efficient algorithms are required to optimize the parameters of the trial wave functions. In Ref.~\cite{Sun:2022frr} it was demonstrated that differential programming and state-of-the-art machine-learning optimizers can be powerful in this setting, outperforming other approaches such as the Stochastic Variational Method (SVM)~\cite{Varga:1995dm} in the context of pionless EFT for few-nucleon systems.  This section outlines the relevant features of the optimization procedure; further implementation details are as in Ref.~\cite{Sun:2022frr}. 

The optimization procedure followed in this work targets the lowest eigenstate of a particular Hamiltonian.\footnote{Other optimization approaches that directly target excited states could be implemented, but are beyond the scope of the present work.} Thus, the nucleon mass, the volume of the box and the values of the LECs are fixed at the beginning of the process. Then, given a wave function ansatz of $N_g$ correlated Gaussians as in 
\Cref{eq:lincombansatz}, the parameters to be optimized are the set of real-valued $c_i$, and the parameters in the correlated Gaussians $\ket{\Psi_i}$, denoted generically as $\theta$. The energy of the trial wave function is given by 
\begin{equation}
\mathcal{E}\left[\Psi(\theta)\right]=\frac{\mathbf{c} \cdot\left(\mathbb{K}^{(2)}+\mathbb{V}^\text{LO}_2+\mathbb{V}^\text{NLO}_2\right) \cdot \mathbf{c}}{\mathbf{c} \cdot \mathbb{N} \cdot \mathbf{c}},
\label{eq:loss}
\end{equation}
where the chosen values for LECs are implicit in the matrix elements of the LO and NLO potentials. The value of $\mathcal{E}\left[\Psi(\theta)\right]$ provides a bound for the ground-state energy, and is minimized through the optimization process, i.e., \Cref{eq:loss} is the ``loss function''. The relevant matrix elements are computed using the expressions in \Cref{app:matrix}, and the sums present in \Cref{eq:glambdaFV,eq:psiLalpha} are evaluated up to $q_\text{cut} = 50$, and $\mathbf{b}_\text{cut} = (50,50)$.

\begin{figure}[t]
\includegraphics[width=\linewidth]{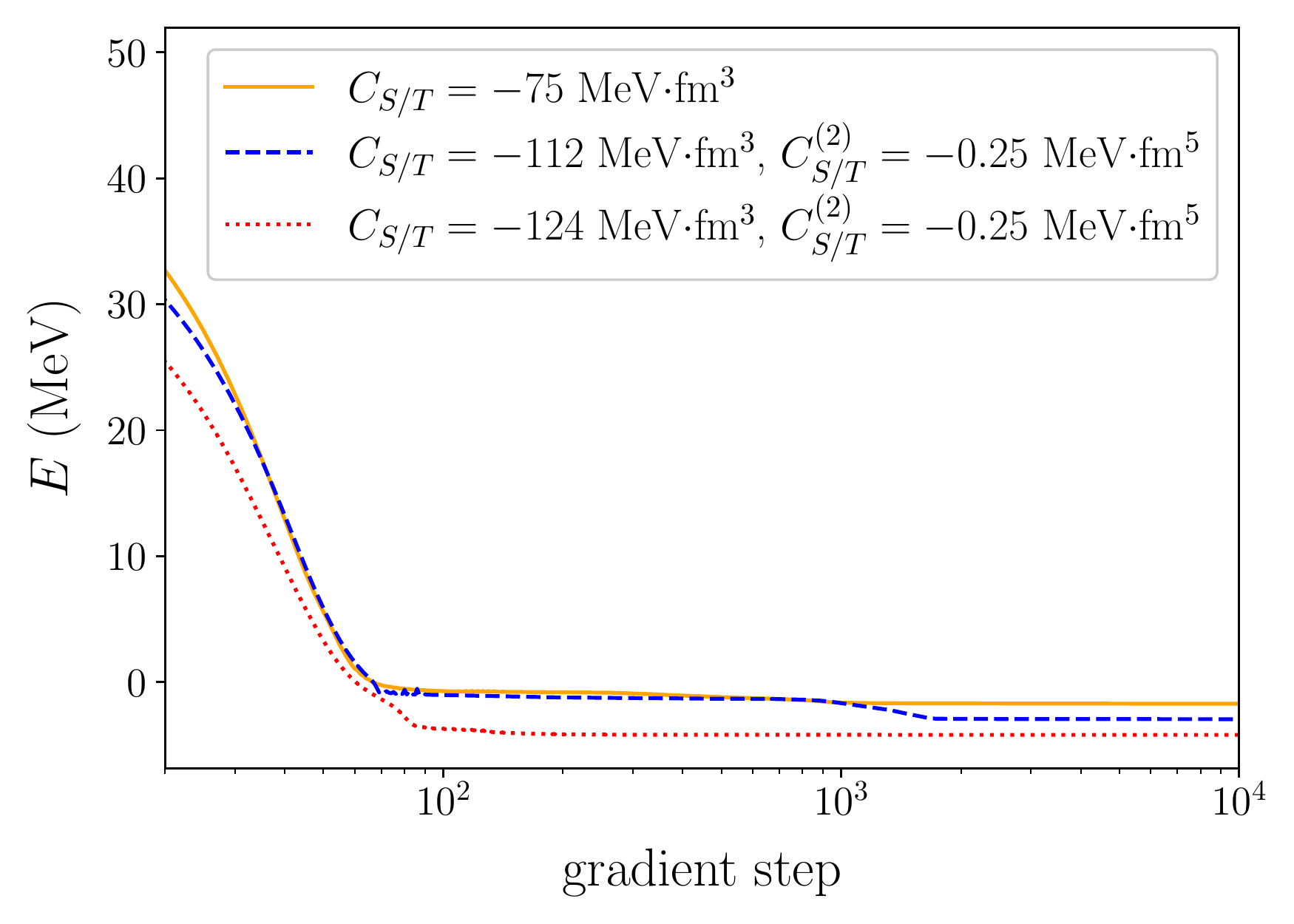}
\caption{Examples of optimization of wave function ans\"atze with $N_g=8$ correlated Gaussians. Each of the three sets of wave functions is optimized during $10^4$ gradient steps and for different LECs, as displayed in the legend.
}
\label{fig:training}
\end{figure}

The same algorithm as in Ref.~\cite{Sun:2022frr} is used to minimize the loss. Automatic differentiation is utilized to compute the gradients with respect to the parameters, with a self-adaptive learning rate (see Appendix B in Ref.~\cite{Sun:2022frr}). The latter is required, since during optimization there are almost-flat directions that require very large learning rates. Other minimizers, such as Adam~\cite{2014arXiv1412.6980K}, do not obviously perform better. Three examples of optimization curves are shown in \Cref{fig:training}. In one case, a second dip of the loss appears, which is a consequence of an almost-flat direction. 

In this work, the evaluation of the energy as the primary optimization target only includes the terms in the Hamiltonian corresponding to $\mathbb{K}^{(2)}$, $\mathbb{V}^\text{LO}_2$ and $\mathbb{V}^\text{NLO}_2$. Once the wave functions have been optimized,
the parameters of the ans\"atze are stored and it is possible to solve the GEVP, as described in \Cref{eq:GEVP}. In the GEVP, the effect of $\mathbb{K}^{(4)}$ and $\mathbb{V}^{sd}_2$ are also considered, but they are excluded from the optimization, since obtaining the gradients for the optimizer in these terms is numerically demanding.
Moreover, $\mathbb{V}^{sd}_2$ in the case of the deuteron channel mixes different spin components, and so, one would require a more complex set of wave functions. 
Including these effects in this manner is sufficient, since they are expected to be perturbative.

Finally, as demonstrated in Refs.~\cite{Detmold:2021oro,Sun:2022frr}, it is useful to combine sets of Gaussians that have been optimized independently, either with the same or different values of the LECs, when solving the GEVP. Indeed, in the application described in \Cref{sec:appLQCD} this leads to improved bounds on both the ground state, and excited states. In practice, to scan over the parameter space or to perform fits, it is more useful to combine sets at different values of the LECs.

\subsubsection{Finite-volume symmetries}

In a finite cubic volume, rotational invariance is broken to a discrete subgroup of transformations, the so-called cubic octahedral group or its double-cover.  Consequently, finite-volume LQCD states enter in irreducible representations (irreps) of this reduced symmetry group rather than as states of well-defined angular momentum (irreps of the rotation group). In addition, the preserved symmetry may be further reduced to the little group corresponding to conserved vectors such as the total three-momentum.
To perform the appropriate matching, these symmetries must also be considered in the FVEFT.

The wave function ansatz in \Cref{eq:ansatz} is in general not covariant under cubic-group transformations. This symmetry can be easily imposed by performing a projection over the appropriate symmetry group to arrive at a wave function ansatz with the desired transformation properties:
\begin{equation}
 \ket{\Psi^\Gamma} =  \sum_{R_i \in \text{LG}(\mathbf P)} \zeta^\Gamma(R_i)  \,\Psi_{L,i}^{\mathrm{sym}}(R_i \mathbf x)\ket{D_S(R_i) \chi_h} ,
 \label{eq:proj}
\end{equation}
 where LG$(\mathbf P)$ stands for ``little group'' and describes the symmetry group in finite volume in a system with certain total momentum $\mathbf P$,  $\zeta^\Gamma(R_i)$ is the character of the rotation $R_i$ in the irrep $\Gamma$, and $D_S(R_i)$ is the representation of the rotation $R_i$ corresponding to spin $S$. In the rest frame, the little group is the octahedral group. For nonzero total momentum, it is the subgroup of the octahedral group that leaves the momentum invariant. The set of irreps, rotations, and characters in the octahedral group and little groups can be found, e.g., in Refs.~\cite{Gockeler:2012yj,Morningstar:2013bda,dresselhaus2007group}.

 This work focuses primarily on states in the rest frame, and the notation for irreps of Ref.~\cite{Morningstar:2013bda} is used. For the dineutron case, which is a spin singlet, the $A_{1g}$ irrep is considered. This irrep couples dominantly to $s$-wave interactions, and corresponds to states that  are fully symmetric under any transformation in the symmetry group (the trivial irrep). 
 As a result, the matrix elements of the Hamiltonian for those states depend only on $C_T$ and $C_T^{(2)}$ at NLO.
 This irrep thus can be used to constrain these LECs. Notice that the rotation of the spin states is trivial in this case. 

In the deuteron case, the $T_{1g}$ irrep is considered, which couples to interactions in channels with total angular momentum and parity $J^P=1^+, 3^+,...$ that are isosinglet. This means that energy levels in this irrep are affected by both $s$- and $d$-wave interactions, and can be used to constrain the LECs $C_S, C_S^{(2)}$ and $C^{(sd)}$.  Two ways of obtaining $T_{1g}$ wave functions are used in this work. First, one can construct a fully symmetric spatial wave function multiplied by a generic deuteron spin-isospin wave function as in \Cref{eq:spiniso}. Second, arbitrary space-spin-isospin wave functions can be projected following \Cref{eq:proj}, for which one needs explicit representation of rotations acting on the spin $D_S(R_i)$, i.e.  Wigner-D matrices for an axial vector. The former approach will be used during optimization and to constrain $s$-wave LECs, and the latter will be used to bound the magnitude of the $s$-$d$-wave mixing term.

\subsection{Numerical examples}
\label{subsec:numerical}

In this section, numerical examples of the NLO effects in the FVEFT method are explored, demonstrating how the different terms in the Hamiltonian shift the energy levels.
In particular, examples are chosen to illustrate the new features explored in this work: irrep projections, relativistic corrections, and NLO terms in the potential. It is also demonstrated how optimizing for the ground state for a sufficiently large set of correlated Gaussians leads to approximately optimized results for low-energy excited states.

For these examples, the physical value of the nucleon and pion masses are chosen, with a box of size $M_\pi L = 4 $. Specifically,
\begin{equation}
    L=5.67 \text{ fm}, \quad M_N L = 27.05\,.
\end{equation}
For the regulator in the interaction potential, the scale is set as $r_0 = 0.2 $~fm.

\subsubsection{Excited states and irreps}

\begin{figure}[t]
\includegraphics[width=\linewidth]{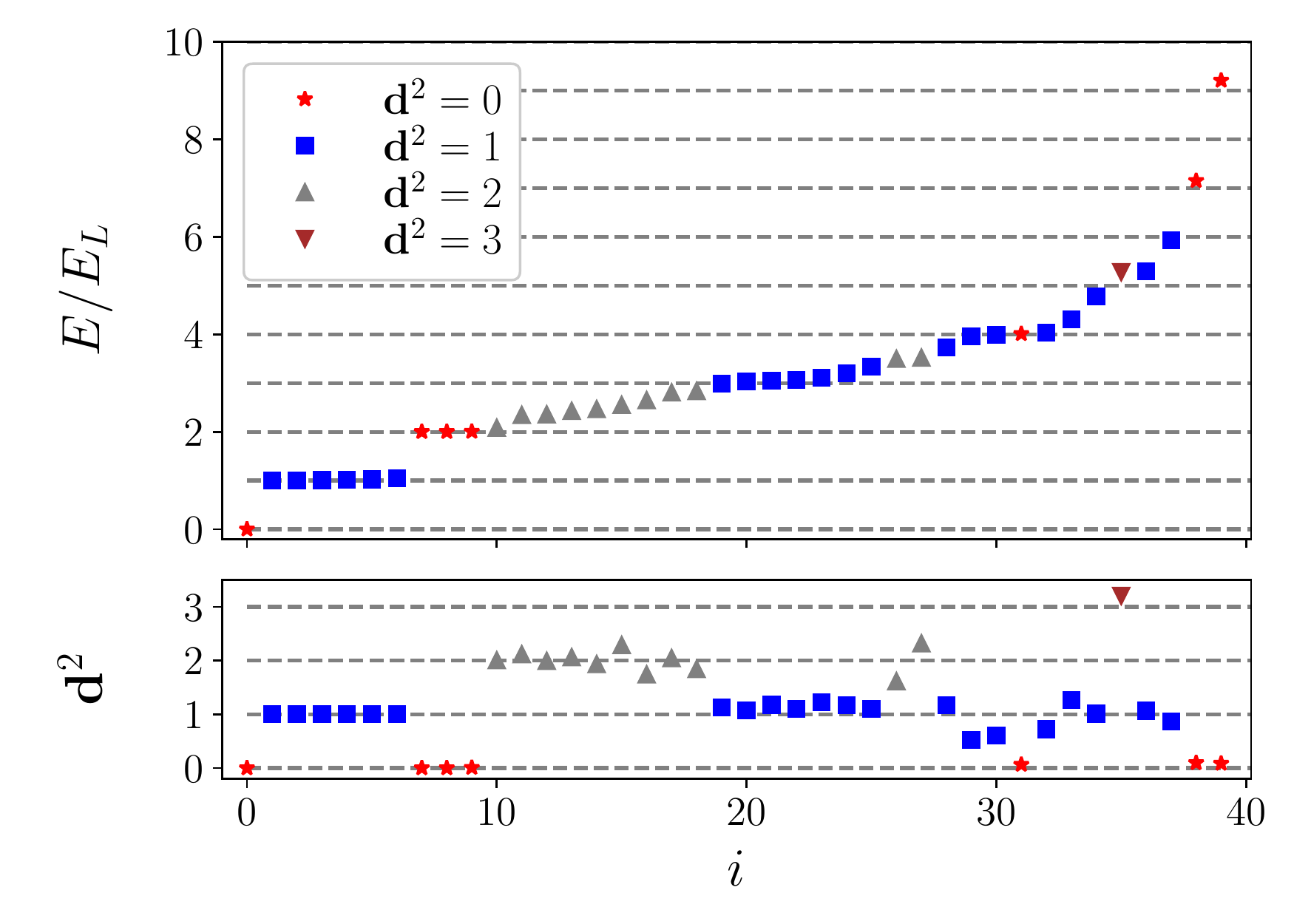}
\caption{Optimized energy bounds in the non-interacting theory in finite volume for several states. No spatial symmetry is assumed for the wave functions. The upper panel shows the energy bounds organized by increasing energy from left to right in units of $E_L$ as in \Cref{eq:free2}. The lower panel shows the calculated value of the squared total momentum of the system in units of $(2\pi/L)^2$. The shape and color of the markers are also used to indicate the total momentum.  These bounds have been obtained by combining 5 different sets of $N_g=8$ correlated Gaussians in a GEVP. They have been optimized for about $2000$ gradient steps at different values of the LO coupling: ${C_S = 0, \pm 25, \pm 50 \text{ MeV}\cdot \text{fm}^3}$ (all other LECs are set to zero). In the combined GEVP, $C_S = 0$  is used.   }
\label{fig:free_nosym}
\end{figure}

A clear illustration of the approach is provided in the free theory, with all LECs set to zero, and in the nonrelativistic limit i.e., with $\left[\mathbb{K}^{(4)}\right]_{ij}=0$. For the dineutron, the overall spin is zero and the behavior of the spin part of the wave function is trivial. The solutions in the free theory are exactly known. The momentum of each particle is quantized by the periodic boundaries as $\mathbf p_i = (2\pi/L) \mathbf n_i $ with $\mathbf{n} \in \mathbb Z^3$, and so the energy is:
\begin{equation}
    E = E_L \left( \mathbf n_1^2 +  \mathbf n_2^2 \right), \quad E_L =  \frac{2\pi^2 }{M_NL^2}. \label{eq:free2}
\end{equation}
Note that several different values of $ \mathbf n_i$ result in degenerate energies. In addition, each level can be assigned to a frame labelled by $\mathbf P^2 = (2\pi/L)^2 \mathbf d^2$, with $\mathbf d =  \mathbf n_1+ \mathbf n_2$. For instance, $E = E_L$ has a 6-fold degeneracy for a symmetric spatial wave function, corresponding to $\mathbf n_1^2=1$ and $\mathbf n_2^2=0$ (and the symmetric combination), where all states are in a frame with overall momentum $\mathbf d^2 = 1$. Moreover, $E = 2 E_L$ has a 33-fold degeneracy, corresponding to (i) 12 states with $\mathbf n_1^2=2$ and $\mathbf n_2^2=0$ (and the symmetric contribution) and $\mathbf d^2 = 2$, (ii) 6 states with $\mathbf n_1^2=\mathbf n_2^2=1$ and $\mathbf d^2 = 4$, (iii) 3 states with $\mathbf n_1^2=\mathbf n_2^2=1$ and $\mathbf d^2 = 0$, and (iv)  12 states with $\mathbf n_1^2=\mathbf n_2^2=1$ and $\mathbf d^2 = 2$.

A numerical example of variational bounds on energies in this system obtained by the approach of \Cref{sec:FVEFT} is shown in \Cref{fig:free_nosym}. The results have been obtained by optimizing 5 sets of correlated Gaussian ans\"atze with $N_g=8$ at different values of the LECs as specified in the figure caption, and then performing a combined GEVP with all LECs set to zero. The upper panel shows the values for the energies in units of $E_L$, and the lower panel shows the overall momentum $\mathbf d^2$ in units of $(2\pi/L)^2$. It can be seen that the first levels with $E=0$ and $E=E_L$ are correctly reproduced, including their degeneracies. In contrast, only a subset of the expected bounds for states with true energies $E=2E_L$ are close to that value. The values of $\mathbf d^2$ for these states differ from integers because the variational wave functions are contaminated by other states. 
One could further improve on these bounds by considering larger sets of optimized wave functions.

\begin{figure}[t]
\includegraphics[width=\linewidth]{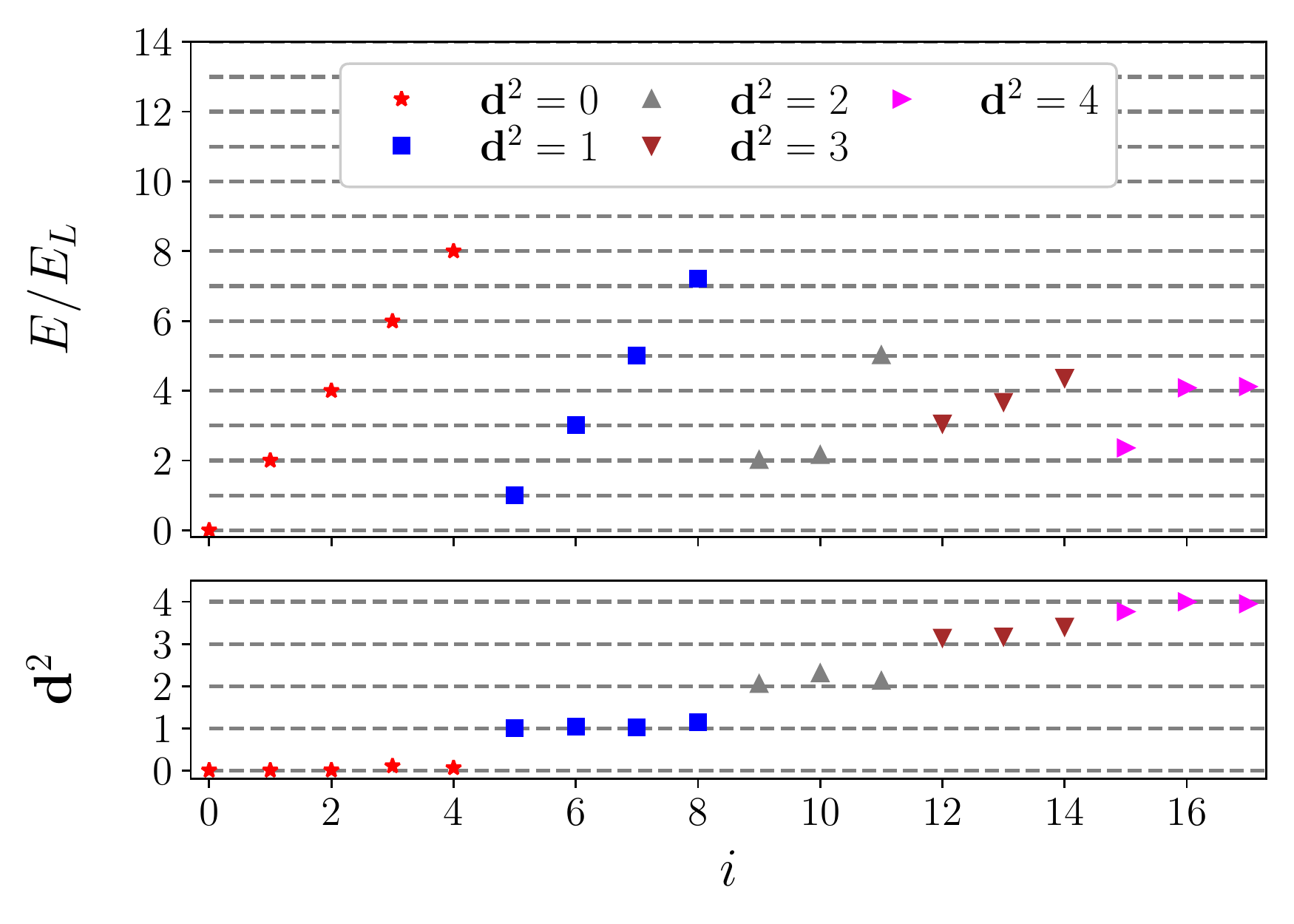}
\caption{Optimized energy bounds in the non-interacting theory in finite volume for several states imposing that the wave function is fully symmetric under permutations and inversions. The energy bounds are organized from left to right by increasing squared total momentum, and with increasing energy at the same momentum. Other notations are as in \Cref{fig:free_nosym}.
These bounds have been obtained by combining in a GEVP three sets of $N_g=8$ correlated Gaussians, optimized at ${C_S = 0.0, \pm 50 \text{ MeV}\cdot \text{fm}^3}$.}
\label{fig:freeA1}
\end{figure}

A second example is provided in \Cref{fig:freeA1}, where the wave functions have been projected to be completely symmetric under permutations and inversions. This corresponds to the $A_{1g}$ irrep for $\mathbf d^2 = 0$, while for $\mathbf d^2 >0$ is equivalent to projecting to the trivial irrep of the little group of the moving frame. Here, 3 sets of correlated Gaussian wave functions with $N_g=8$ are optimized at different values of the LECs as specified in the figure caption. In this case, the bounds are much closer to the known free energy levels than in the example of \Cref{fig:free_nosym}: the correct degeneracies for the levels emerge at $E=2E_L$, and those at $E=3E_L$ are also partially reproduced. It is also interesting to note that the first few levels with $\mathbf{d}^2=0$ and $\mathbf{d}^2=1$ are very well reproduced, but starting from  $\mathbf{d}^2=2$, significant deviation can be seen. Larger sets of trial wave functions would also allow for further improvements in the fidelity of the spectrum.

\subsubsection{Relativistic corrections}

Next, the effect of the $O(p^4)$ relativistic correction in the spectrum is explored, once again in the setting of the free theory, for which the answer is exactly known. In this case, the relativistic correction is expected to shift the energy as
\begin{equation}
    \Delta E^\text{(4)} = - \frac{E_L^2}{8 M_N} \left( \mathbf n_1^4 +  \mathbf n_2^4 \right). \label{eq:deltaE4}
\end{equation}
Note that this breaks the degeneracy between some levels. For instance, the 33 energy levels with $E=2E_L$ in the nonrelativistic limit become two different sets of levels, 12 levels with $\Delta E^\text{(4)} = -E_L^2/(2M_N)$ and 21 levels with $\Delta E^\text{(4)} = -E_L^2/(4M_N)$.

\begin{figure}[t!]
\includegraphics[width=\linewidth]{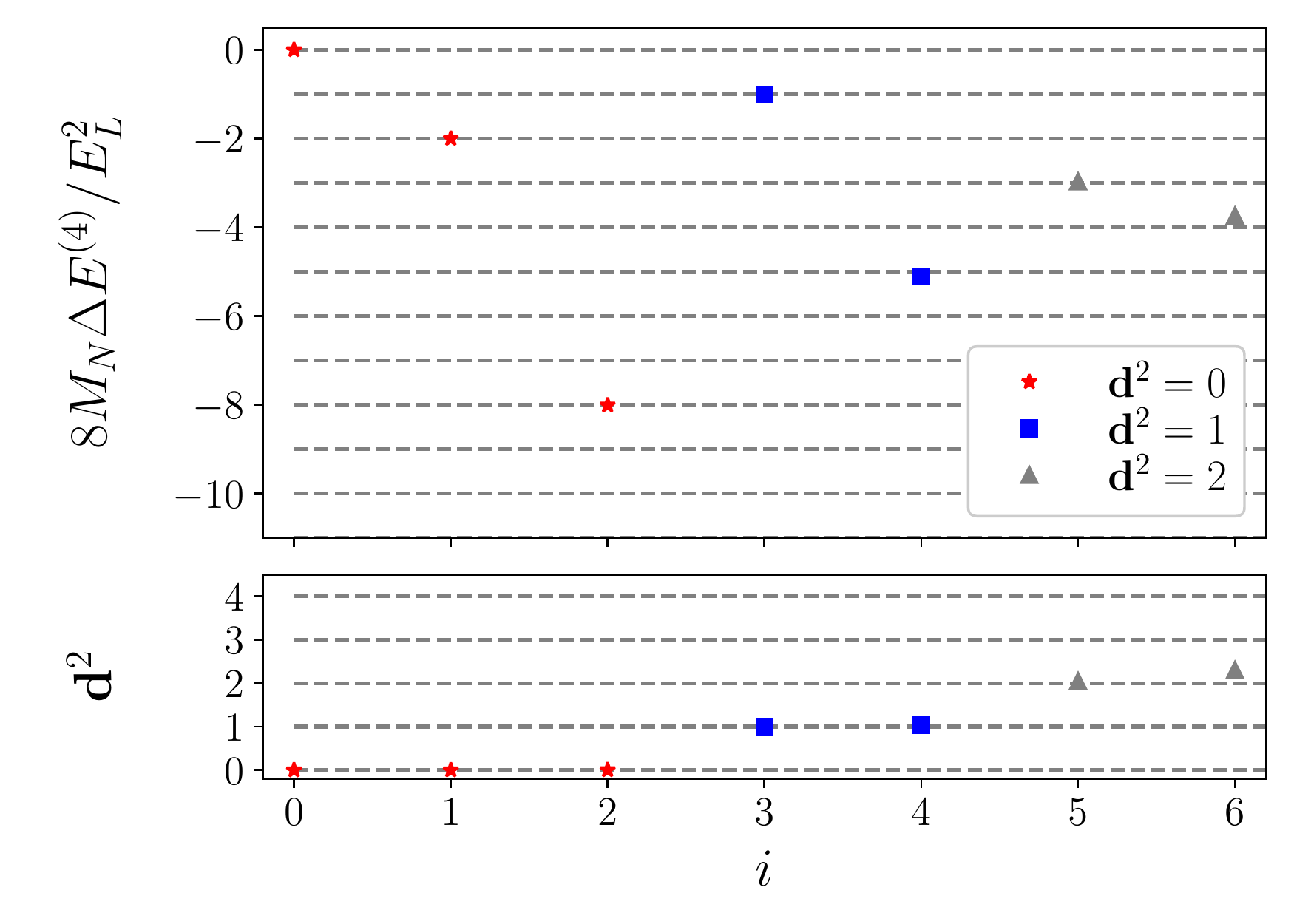}
\caption{ Shifts in the variational energy bounds due to the first relativistic correction. The $y$-axis is given in dimensionless units such that the exact predictions for the shifts are integers, see \Cref{eq:deltaE4}. All notation is as in \Cref{fig:freeA1}, and the same trial wave functions as used for the results in that figure have been used.}
\label{fig:rel4}
\end{figure}

An example of the effect of the relativistic correction is shown in \Cref{fig:rel4}. Here, the same set of correlated Gaussians are used as in the results shown in \Cref{fig:freeA1}, a projection to the fully symmetric wave function is performed, and the relativistic correction $\left[\mathbb{K}^{(4)}\right]_{ij}$ is included as a part of the GEVP.\footnote{The relativistic correction could also be included during optimization, but its effect is small and perturbative, and it is sufficient to include it at the level of the GEVP.} The figure shows the shift of the lowest few energy levels for $\mathbf d^2=0,1,2$ when adding the relativistic term. This shift is correctly reproduced to better than relative $2\%$ accuracy for the lowest levels with $\mathbf d^2=0$ and $1$. In contrast,  for $\mathbf d^2=2$ some discrepancies can be seen. In particular, the shifts should be $-2$ and $-4$ in the units of the $y$-axis of the figure, and values around $-3$ and $-4$ are found. This indicates that the optimized wave functions are contaminated by higher-lying states, which can also be seen in the fact that the state with $i=6$ has a value of $\mathbf{d}^2$ that is larger than 2. 

\subsubsection{Effect of the $s$-$d$-wave mixing term}

A final numerical example is given for the deuteron channel, where the $s$-$d$-wave mixing term in $V_2^{sd}$ contributes to the spectrum in the $T_{1g}$ irrep.\footnote{See Ref.~\cite{Briceno:2013bda} for a study of the effect of this term based on the two-particle quantization condition.} At second order in perturbation theory, the effect of this term on the ground state is expected to be
\begin{equation}
    \Delta E^{sd} = E - E \big \rvert_{C^{(sd)}=0} \propto (C^{(sd)})^2 + \mathcal O\left[(C^{(sd)})^4\right],
\end{equation}
where $ \Delta E^{sd}$ is the shift due to the $s$-$d$-wave mixing term. For higher excited states, other effects are possible. In particular, in the free theory the state with $\mathbf d^2=0$ and $\mathbf n_1=\mathbf n_2=1$ in the $T_{1g}$ irrep has a 6-fold degeneracy: $A_{1g}$ irrep in the spatial wave function combined with $T_{1g}$ in spin, and $E_g$ in the spatial part combined with $T_{1g}$ in spin. These states are degenerate in the free theory. If only the $V_2^{sd}$ term is added to the Hamiltonian, perturbation theory on degenerate states can be used to show that these levels acquire a shift proportional to $|C^{(sd)}|$. In contrast, if any other term in the potential, e.g.~$V^\text{LO}$, is different from zero, these states will not be degenerate at zero $s$-$d$-wave mixing and the effect of the LEC will be quadratic. 

\begin{figure}[t!]
\includegraphics[width=\linewidth]{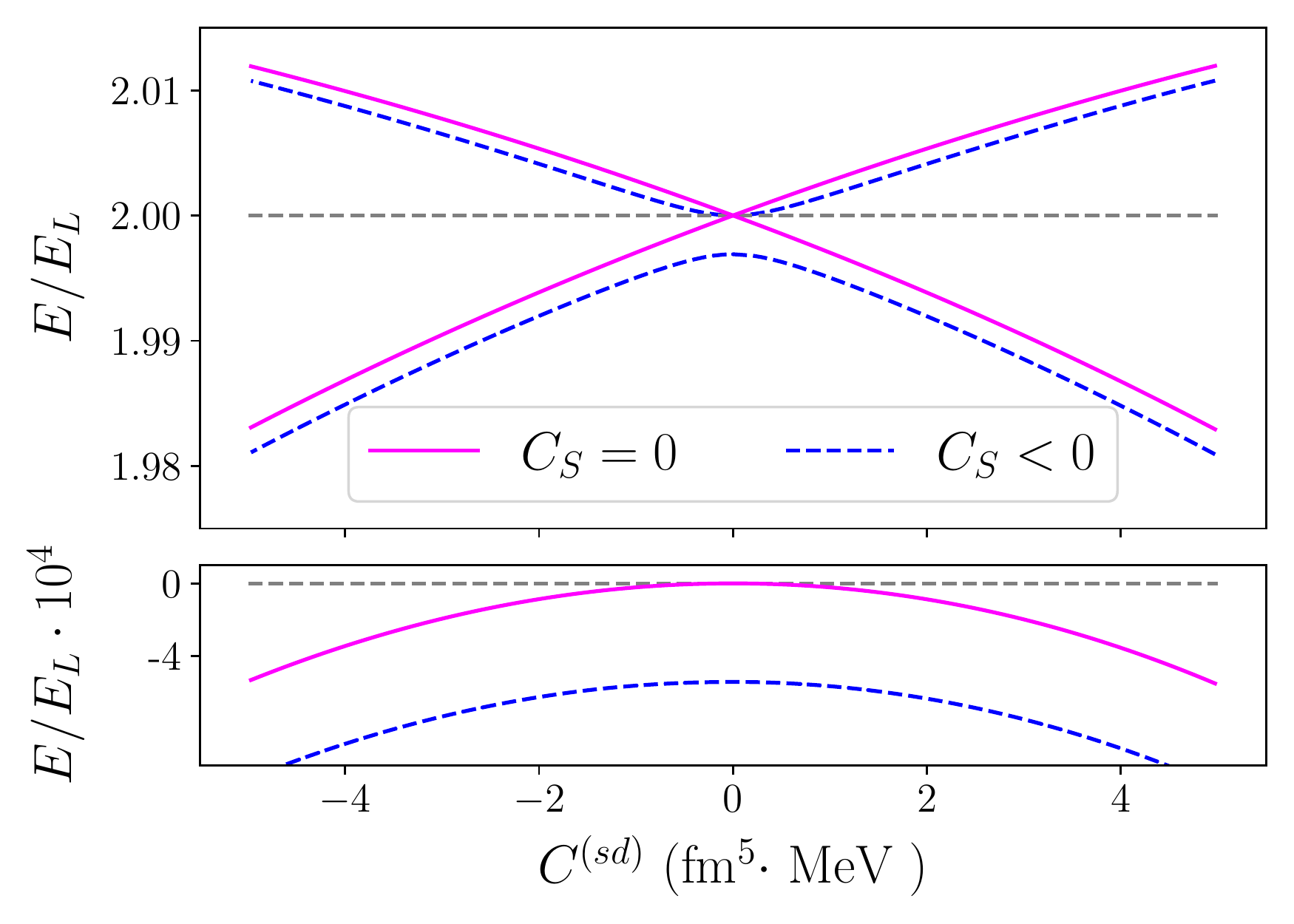}
\caption{ Dependence of energy bounds of the ground and excited states in the deuteron channel on the $s$-$d$-wave mixing LEC $C^{(sd)}$. The lower panel corresponds to the ground state, and the upper panel to the next excited states. The dashed gray lines indicate exact solutions in the free theory. The solid magenta lines are obtained by setting all LECs to zero except $C^{(sd)}$, and the dashed blue lines by setting $C_S=-2.5$ fm$^3 \cdot$ MeV and varying $C^{(sd)}$. The same five sets of $N_g=8$ correlated Gaussians as in \Cref{fig:free_nosym} are used.  }
\label{fig:sd}
\end{figure}

To evaluate the effect of this term, 5 sets of $N_g=8$ correlated Gaussians are optimized. Three different third-component spin projections are applied to each set resulting in a $120 \times 120$ matrix in the GEVP.
The dependence of the energy bounds on $C^{(sd)}$ is displayed in \Cref{fig:sd}. The linear and quadratic responses seen are as expected from the arguments above.

\section{Constraints on LECs from lattice QCD}
\label{sec:appLQCD}

\begin{table*}[th!]
\centering
\begin{tabular}{|c|c|c|c|c|}
\hline
Channel                   & Irrep                           & \multicolumn{3}{c|}{Energy shifts (MeV)} \\ \hline
\multirow{1}{*}{deuteron} & $T_{1g} = A_{1g} \otimes T_{1g}$ & -3.4(7)   & 29.5(1.3)   & 71.5(1.4) \\ \cline{1-5}
dineutron                 & $A_{1g}$                        &  -3.3(7) & 30.7(1.2)   & 72.6(1.4)  \\ \hline
\end{tabular}
\caption{GEVP energy bounds from Ref.~\cite{Amarasinghe:2021lqa} used in this work. ``Energy shifts'' refers to the difference between the LQCD energy bounds and the two-nucleon threshold. 
}
\label{tab:spectrum}
\end{table*}

This section presents a FVEFT$_{\pislash}$ analysis of the spectra of Ref.~\cite{Amarasinghe:2021lqa} for the dineutron and deuteron channels. The goal is to obtain constraints on the values of the LECs, $C_{S/T}, C^{(2)}_{S/T}$ and $C^{(sd)}$. In the LQCD calculation of Ref.~\cite{Amarasinghe:2021lqa}, the lattice spacing and the nucleon mass are 
\begin{equation}
    a = 0.1453(16) \text{ fm}, \quad aM_N = 1.20467(57),
\end{equation}
and the spatial lattice extent is $L/a = 32$. Since the uncertainty in both the lattice spacing and the nucleon mass are much smaller than those in the two-nucleon energies, they are neglected in this analysis. The energy bounds used as constraints are given in \Cref{tab:spectrum}. In the deuteron channels, the levels in the  $T_{1g}$ irrep that are dominated by $s$-wave interactions are used. In the dineutron channel, the $A_{1g}$ irrep is used. Only the energy bounds below the $t$-channel cut and dominated by baryon-baryon operators of Ref.~\cite{Amarasinghe:2021lqa} are used here. States above the $t$-channel cut are outside of the range of validity of pionless EFT. The GEVP states in Ref.~\cite{Amarasinghe:2021lqa} that are dominated by hexaquark operators are not used in the analysis of this work, as new degrees of freedom would be needed to describe them in the EFT.
It is important to note that the LQCD results in Ref.~\cite{Amarasinghe:2021lqa} are themselves variational upper bounds on the eigenvalues of the system, rather than estimates of the corresponding energy levels. If the true eigenenergies on the system were significantly below these bounds, the extracted LECs would change.

A variational analysis is undertaken as detailed in \Cref{sec:FVEFT}, with five different sets of $N_g=8$ correlated Gaussians ans\"atze optimized at different values of the LO and NLO coupling. The couplings are chosen such that they span the region around the a posteriori fit values. For example, for the scheme set by $r_0 = 0.2$ fm, the LECs are chosen in the range: $C_{S/T} \in [-50, -130]$ fm$^3 \cdot$ MeV for the LO LECs and $C^{(2)}_{S/T} \in [0, -2]$  fm$^5 \cdot$ MeV for the NLO LECs.
In each case, different groups of 5 sets of $N_g=8$ Gaussians optimized independently at different values of the LECs produce equivalent results, with the resulting difference in the energy bounds being at least one order of magnitude smaller than the uncertainties in the LQCD energy bounds in \Cref{tab:spectrum}.

\subsection{Constraints on the LO couplings}
\label{subsec:LO}

\begin{figure*}[ht!]
     \centering
     \subfloat[deuteron channel \label{fig:dLO}]{
     \includegraphics[width=0.5\textwidth]{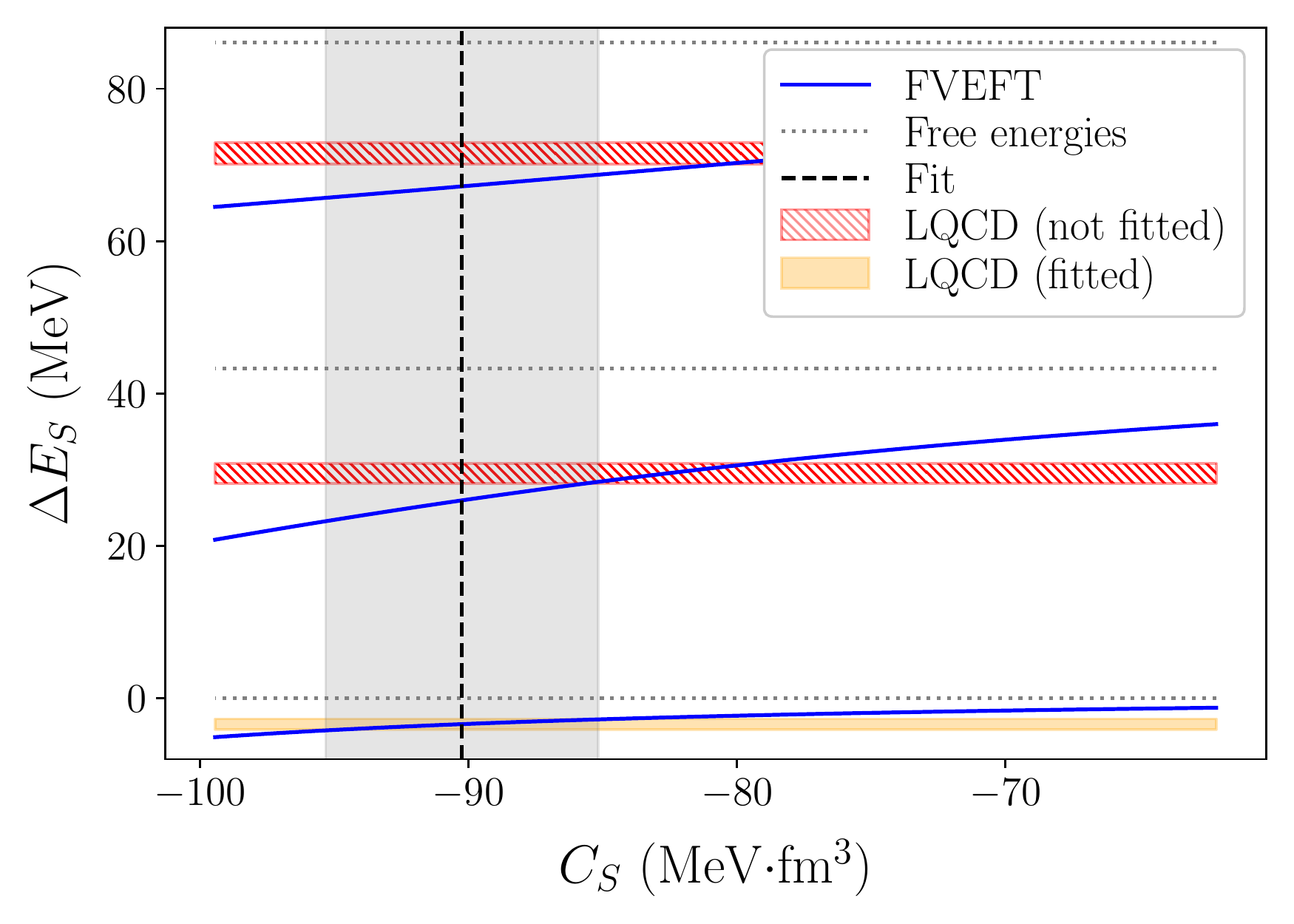}
    }
    \subfloat[dineutron channel\label{fig:nnLO}]{
     \includegraphics[width=0.5\textwidth]{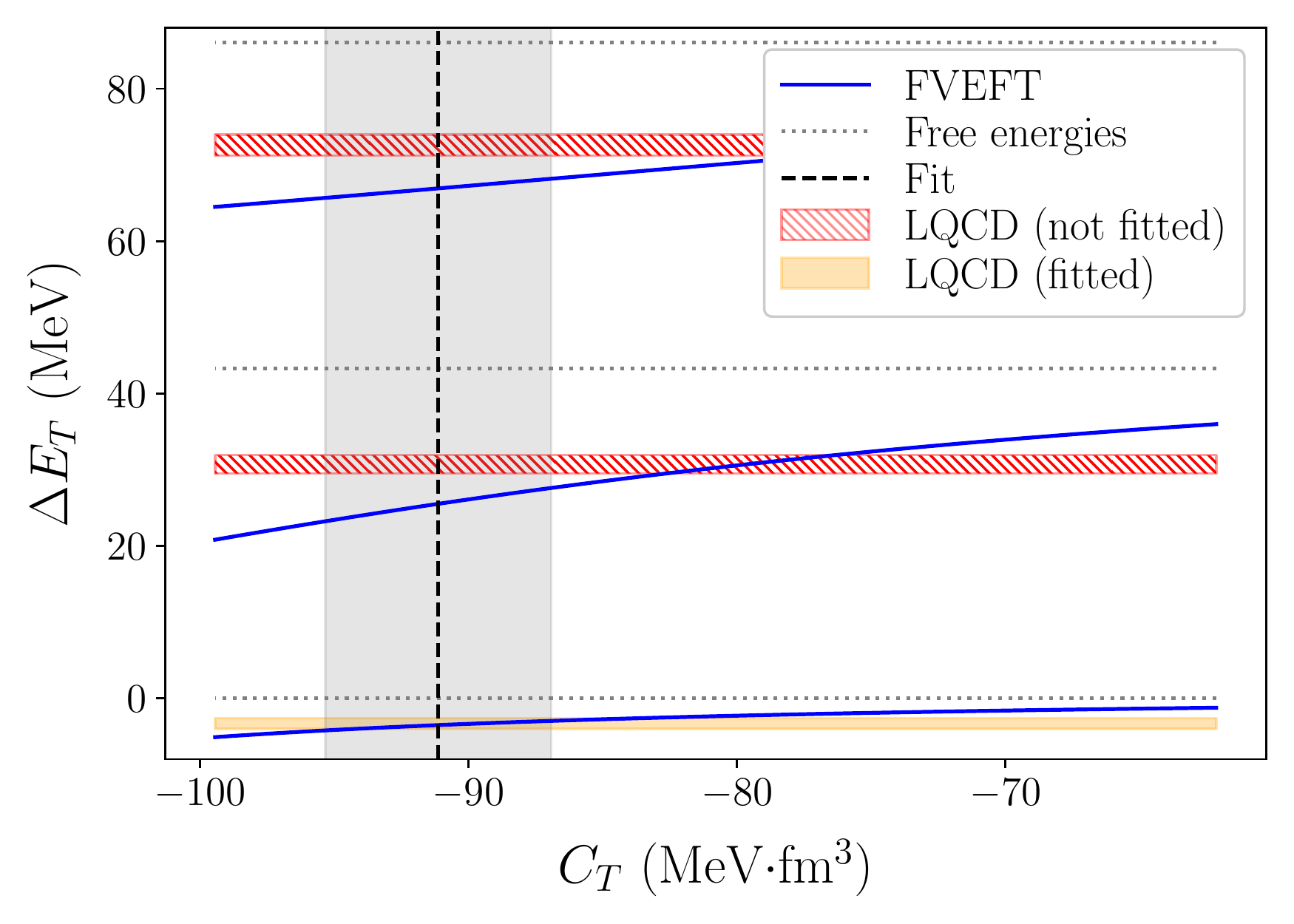}
    }
    \caption{ Comparison between FVEFT$_{\pislash}$ predictions and LQCD data for the finite-volume energies in the deuteron (left) and dineutron (right) channels. The blue line shows the dependence of the variational spectrum predicted by FVEFT$_{\pislash}$ on the LO couplings (the first relativistic correction is also included). Horizontal bands correspond to the energy bounds from Ref.~\cite{Amarasinghe:2021lqa}: the orange band for the level used in the matching, and the red dashed bands for those not included in the fit. Dotted horizontal grey lines correspond to the energies in the free theory (all LECs set to zero). The dashed black line is obtained by matching the ground state of each channel, and the grey band indicates the 1$\sigma$ uncertainty in the LECs (see \Cref{eq:fitLO}). } 
    \label{fig:LO}
\end{figure*}

A first analysis is performed using only the LO effective theory in each isospin channel. Here, the scheme of the LECs is fixed by setting $r_0=0.2$ fm, as in previous work~\cite{Sun:2022frr}. In this case, the analysis will be restricted to the ground state since momentum dependence is expected to be more impactful for excited states.
Given optimized wave function ans\"atze, matching to the ground state LQCD energies constrains the value of the $C_S$ or $C_T$ directly.

This analysis yields effective couplings 
\begin{align}
\begin{split}
     C_S &=  -90(5) \text{ MeV}\cdot\text{fm}^3 , \\ 
     C_T &= -91(4)  \text{ MeV}\cdot\text{fm}^3.   \label{eq:fitLO}
\end{split}
\end{align}
Here, the first relativistic correction is included, although it makes very little difference compared to the uncertainties of the LECs since the state is near threshold.
Since these LECs are very close, spin-dependent interactions are subdominant with respect to spin-independent interactions, as observed in Refs.~\cite{Detmold:2021oro,Sun:2022frr}. The difference in these LECs compared with those obtained in the similar analyses of Refs.~\cite{Barnea:2013uqa,Eliyahu:2019nkz,Detmold:2021oro,Sun:2022frr} is a result of the different LQCD energy spectrum used in this work, namely that of Ref.~\cite{Amarasinghe:2021lqa}, rather than Ref.~\cite{NPLQCD:2012mex}.

Having fixed the effective couplings by matching to the LQCD ground-state energies, one can examine how well excited states are described. This comparison is displayed in \Cref{fig:LO}, where the dependence of the spectrum on the LO LECs for the two isospin channels is compared with the ground- and excited-state energy-bounds from LQCD.  
As can be seen, after fixing the LO LEC by matching to the ground state, excited states are not correctly reproduced, and the disagreement is larger as the energy of the state grows. This indicates that LO EFT$_{\pislash}$ is insufficient to describe the LQCD results, and is the motivation to extend the analysis to use NLO EFT$_{\pislash}$.

\subsection{Constraints on the NLO couplings}
\label{subsec:NLO}

Using the three energy levels in each isospin channel shown in \Cref{tab:spectrum}, the set of LO and NLO LECs, $C_{S/T}$ and $C^{(2)}_{S/T}$ can be simultaneously constrained. Since there are three levels and two parameters, this entails a (correlated) fit with a single degree of freedom. The first relativistic correction is included; excluding it reduces the fit quality by increasing the $\chi^2$ by about one unit, and changes the best fit parameters by over 2 standard deviations. Statistical uncertainties in these fits are propagated from the LQCD energies using the derivative method. Explicitly, the covariance matrix ($V_{nm}$) of the parameters $p_n$ can be obtained as
\begin{equation}
 V_{n m}=\left(\frac{\partial E_i^\text{pred}}{\partial p_n}\left(C^{-1}\right)_{i j} \frac{\partial E_j^\text{pred}}{\partial p_m}\right)^{-1} ,
\end{equation}
where $E_i^\text{pred}$ is the predicted energy level, and $C$ is the covariance matrix of the energies obtained from the data.

\begin{figure*}[ht!]
     \centering
     \subfloat[deuteron channel \label{fig:dNLO}]{
     \includegraphics[width=0.5\textwidth]{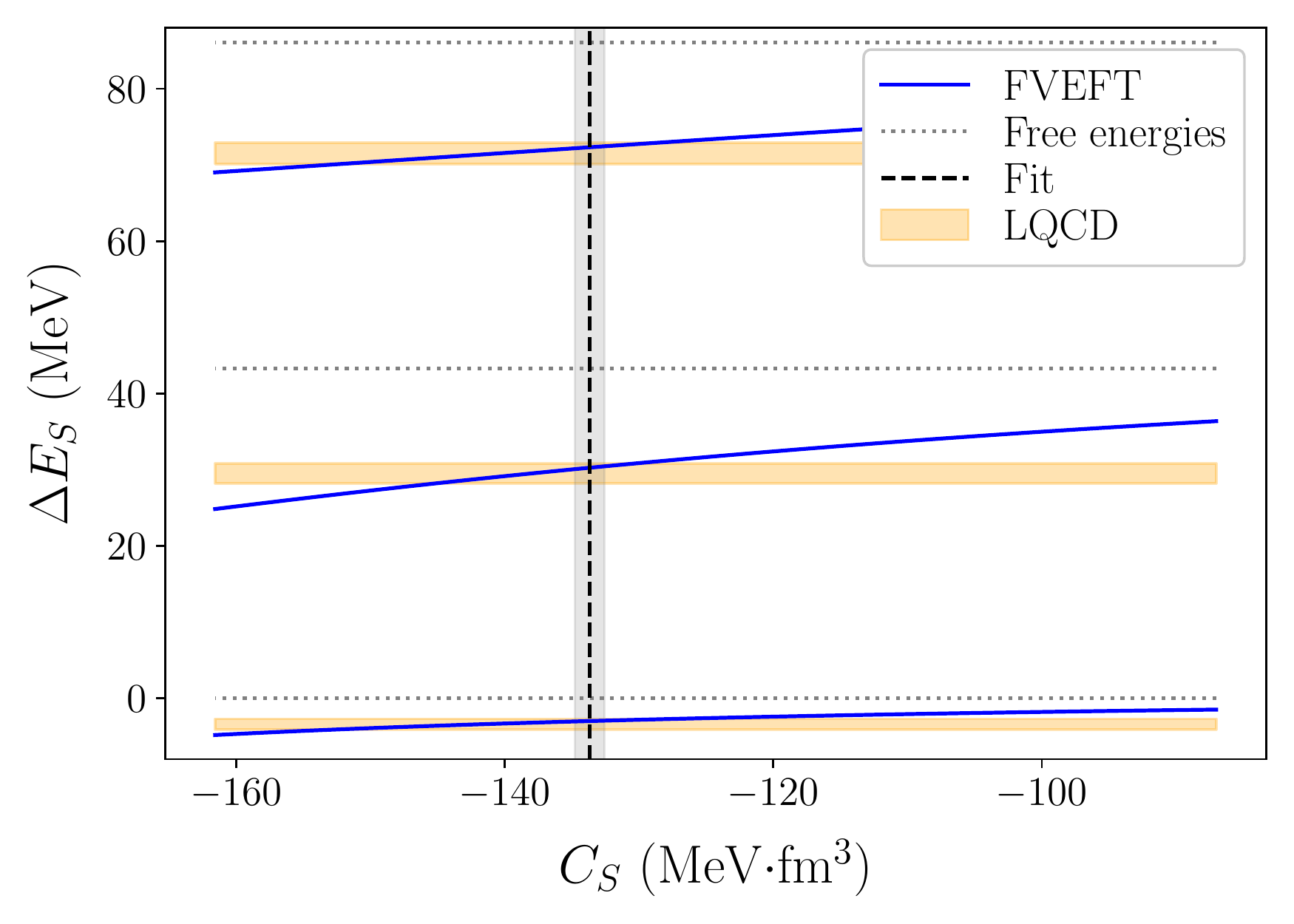}
    }
    \subfloat[dineutron channel\label{fig:nnNLO}]{
     \includegraphics[width=0.5\textwidth]{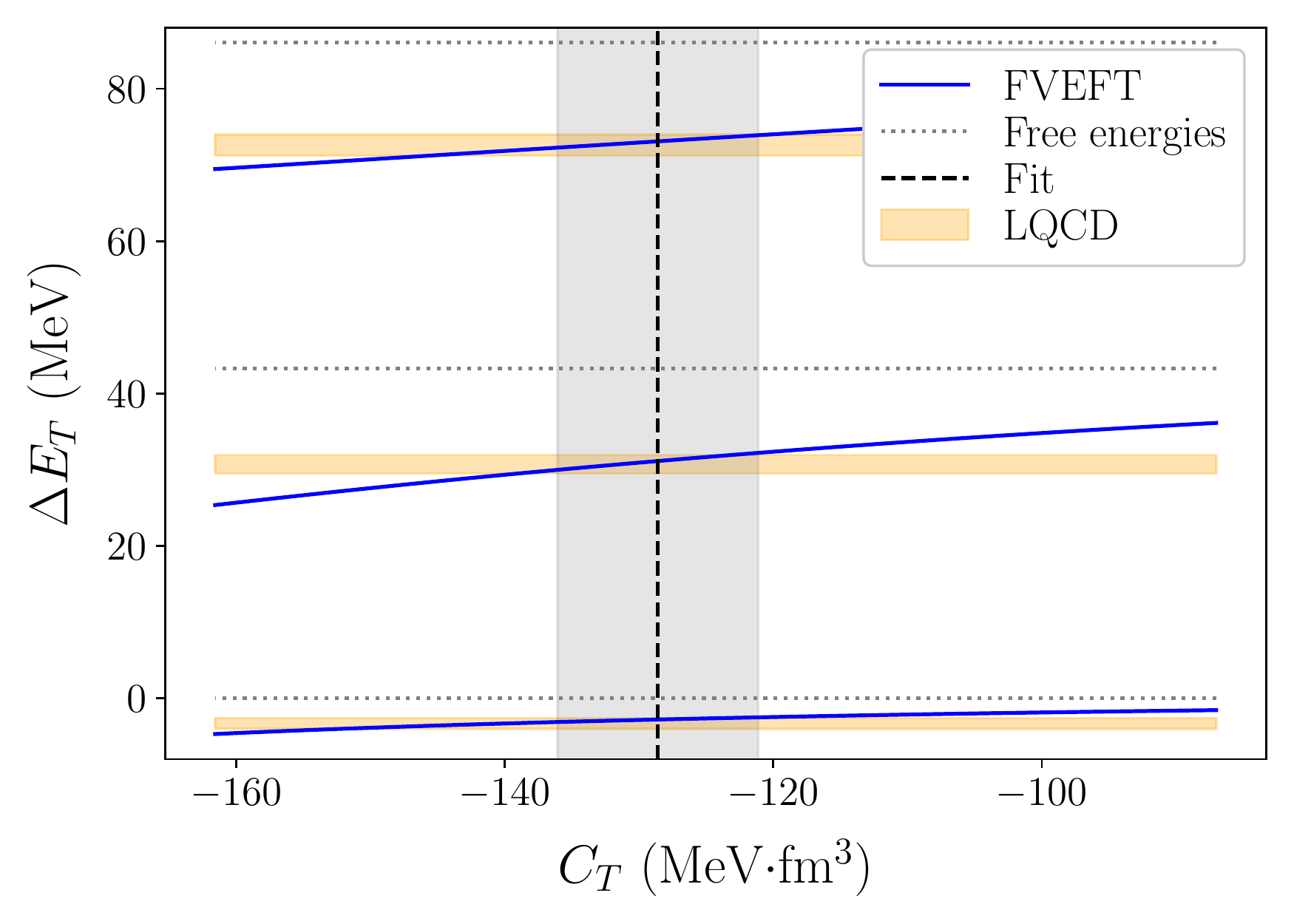}
    }
    \caption{ Visualization of the fits including the LO and NLO couplings at $r_0=0.2$ fm for the deuteron (left) and dineutron (right) channels.  The blue lines show the variational spectrum predicted by FVEFT$_{\pislash}$ as a function of the leading order couplings, while fixing the NLO LEC, $C_{S/T}^{(2)}$ to its best fit value in \Cref{tab:fits}. The first relativistic correction is also included.
     Horizontal orange bands correspond to the energy bounds from Ref.~\cite{Amarasinghe:2021lqa}. Dotted grey lines correspond to the energies in the free theory. The dashed black line indicate the best fit result for the LO coupling, and the grey bands indicate the 1$\sigma$ uncertainty  
     }
    \label{fig:NLOfits}
\end{figure*}

\begin{figure}[ht!]
\includegraphics[width=\linewidth]{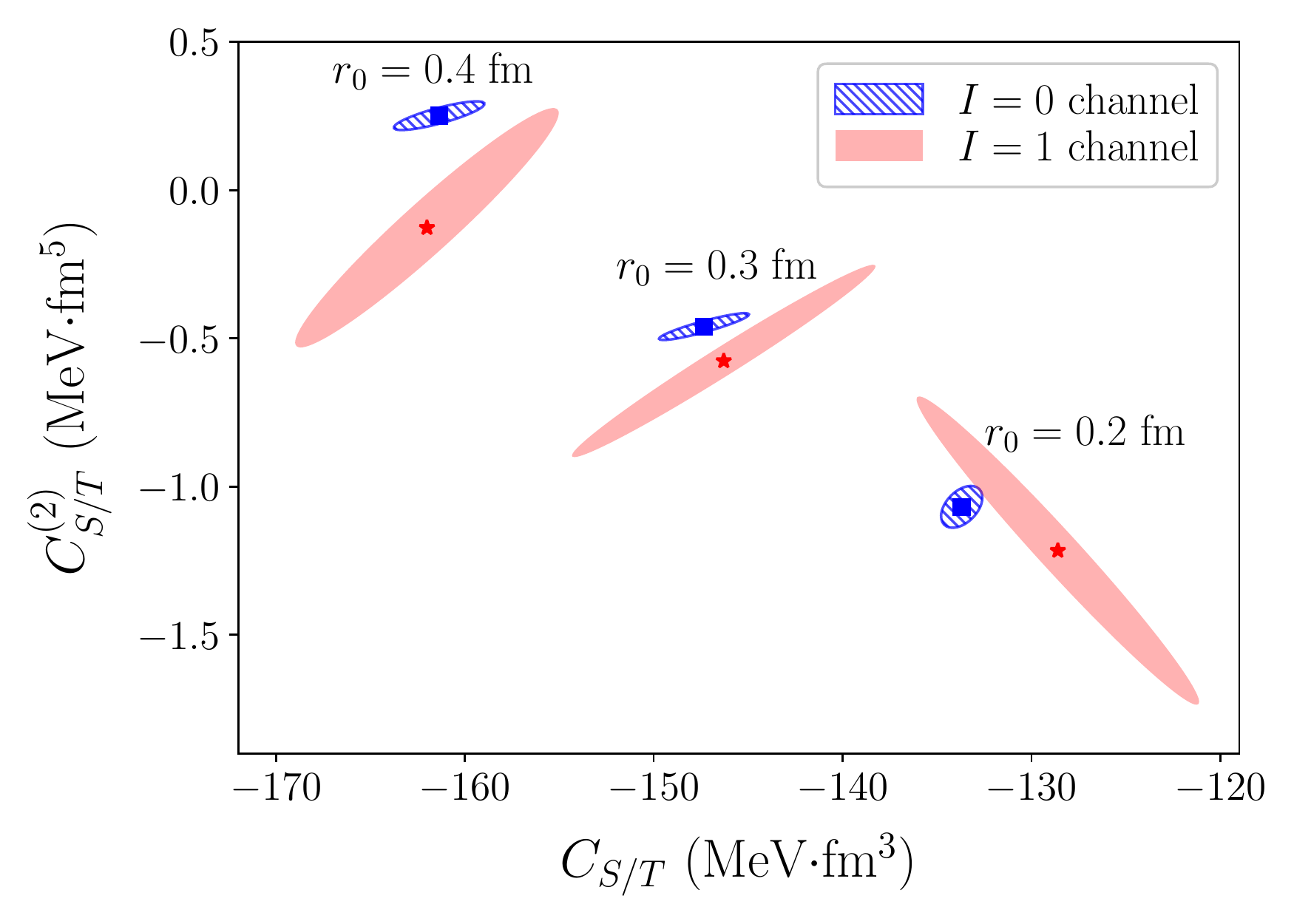}
\caption{ One standard deviation confidence interval for the LO ($C_{S/T}$) and NLO ($C^{(2)}_{S/T}$) couplings obtained after fitting three energy levels in each isospin channel. Results for three different schemes are shown, as indicated by the value of the scale $r_0$ given above each set of ellipses. The ellipses with blue stripes correspond to the deuteron channel, while red ellipses denote the dineutron channel. }
\label{fig:scaledep}
\end{figure}

Fits to both channels using three different regulator schemes for the LECs are performed, i.e., ${r_0 \in \{0.2, 0.3,0.4\}}$ fm. The results for the LECs in both channels are given in \Cref{tab:fits}.
As can be seen from the $\chi^2$, all fits have a good quality. For the case of ${r_0=0.2}$ fm, the values of the LO couplings are significantly different than those obtained in the analysis of the ground state energies only; see \Cref{eq:fitLO}. Moreover, the ground-state energy is equally well reproduced.

Figure~\ref{fig:NLOfits} provides a representation of the fit quality for the case of $r_0=0.2$ fm and both isospin channels. Here, the blue lines correspond to the NLO FVEFT$_{\pislash}$ predictions as a function of the LO LEC, and with the value of the NLO LECs fixed at their best fit result. The agreement between the LQCD energy bounds, and the FVEFT$_{\pislash}$ predictions is very good, as anticipated by the low value of the $\chi^2$ in \Cref{tab:fits}.

A visualization of the constraints on NLO EFT$_{\pislash}$ from the matching is provided in \Cref{fig:scaledep}, where the confidence intervals for the various LECs are shown. Due to the correlations in the LQCD energies, the determinations of the LECs in each channel are also significantly correlated. Interestingly, the LECs in the deuteron channel are significantly better constrained, even though the uncertainties of the LQCD energy levels are approximately the same. This results from the covariance matrix of the LQCD energies being very different, even though the diagonal entries are of similar magnitude.

In considering fits using different FVEFT$_{\pislash}$ schemes (different values $r_0$), some scale dependence can be seen. While most of the effect of changing the scale can be reabsorbed by modifying the values of the LECs, the $\chi^2$ values in \Cref{tab:fits} and the shape of the $1\sigma$ ellipses in \Cref{fig:scaledep} do vary with $r_0$. This residual scale dependence is expected, since the regulator induces mixing between NLO terms and operators with a higher number of derivatives. Nevertheless, in the range of values of $r_0$ used here, the $p$-values of the fits are in the region 0.5 to 0.8.

\begin{table*}[th!]
\begin{tabular}{|c||c|c|c||c|c|c|}
\hline
$r_0$ (fm) & $C_S$ ($\text{MeV}\cdot\text{fm}^3$) & $C^{(2)}_S$ ($\text{MeV}\cdot\text{fm}^5$) & $\chi^2_{I=0}$ & $C_T$ ($\text{MeV}\cdot\text{fm}^3$) & $C^{(2)}_T$ ($\text{MeV}\cdot\text{fm}^5$) & $\chi^2_{I=1}$ \\ \hline
0.2        &   -134(1)    &     -1.07(7)        &  0.34               &  -129(8)     &    -1.2(5)         &     0.55           \\ \hline
0.3        &   -147(2)    &  -0.46(5)           &       0.21         &  -146(8)     &    -0.6(3)        &      0.42          \\ \hline
0.4        &   -161(2)    &    0.25(5)         &    0.05            &   -162(7)    &     -0.1(4)        &     0.43           \\ \hline
\end{tabular}
\caption{Summary of the fit results including LO and NLO LECs for both isospin channels, and using three different schemes for the LECs defined by the value of $r_0$. }
\label{tab:fits}
\end{table*}

\subsection{$s$-$d$-wave mixing in the deuteron}
\label{subsec:sd}

Finally, by including the term $V_2^{sd}$ in the potential in the GEVP of the FVEFT$_{\pislash}$, the $s$-$d$-wave mixing coupling in the deuteron channel can also be constrained from the same set of LQCD energy bounds as used for the analysis in the previous subsection. As the effect of $C^{(sd)}$ on these energy levels is quadratic, only its magnitude can be constrained. Since the fits in \Cref{tab:fits} are of high quality, there is no statistical necessity to include $V_2^{sd}$, and so the effect of this coupling can only be bounded.

Instead of performing a fit with no degrees of freedom to constrain the magnitude of this parameter, the $\chi^2$ function is used to bound its magnitude. Specifically, given the best fit parameters in \Cref{tab:fits} at $r_0=0.2$ fm, and requiring $\chi^2<\chi^2_\text{min}+1$, bounds the $s$-$d$-wave mixing LEC as
\begin{equation}
    |C^{(sd)}| < 0.6\ \text{MeV}\cdot\text{fm}^5.
\end{equation}
As can be seen, the bound is about a factor of 2 smaller than the central value found for $C_S^{(2)}$. 

Alternatively, one could consider using other energy levels from Ref.~\cite{Amarasinghe:2021lqa} to constrain $C^{(sd)}$. For example, those in the $T_{1g}$ irrep that have coupling to $d$-wave interactions, e.g., resulting from states with nontrivial spin-spatial coupling. The bounds on the energies of these states in Ref.~\cite{Amarasinghe:2021lqa} lie below the non-interacting energies. By contrast, the effect of $C^{(sd)}$ on those levels is always repulsive, suggesting that this LEC cannot be the only source of the negative shifts. It is thus clear that it would be necessary to also include $d$-wave interactions in an analysis of these levels. The corresponding operators would contain four derivatives and are beyond the scope of this work.

\section{Summary and Outlook}
\label{sec:conclusion}

This work demonstrates that EFTs formulated in a finite volume can be used to analyze LQCD spectra of two-nucleon systems. In particular, finite-volume pionless EFT has been applied to analyze two-nucleon GEVP energy bounds obtained in Ref.~\cite{Amarasinghe:2021lqa} in the dineutron and deuteron channels, yielding constraints on the LO and NLO LECs of the EFT$_{\pislash}$ Hamiltonian.

This analysis makes use of several developments in the usage of FVEFT$_{\pislash}$. First, it has been shown that it is possible to obtain variational bounds not only on the ground state, but also on several excited states of the interacting theory. Second, projection to irreducible representations of the finite-volume symmetry group has been performed, along with measurements of the total momentum of the system. Combined, this allows excited-state energy levels with definite quantum numbers to be mapped from lattice QCD into variational bounds in the FVEFT. Third, matrix elements of the first relativistic correction to the FVEFT$_{\pislash}$ Hamiltonian have been computed, which is relevant for analysis of excited states; omitting these corrections reduces the fit quality and changes the best-fit values of the LECs by over 2$\sigma$.
 Fourth, the effects of operators with up to two derivatives, the NLO $s$-wave term, and the $s$-$d$-wave mixing term have been studied. The effects of the NLO operator in the Hamiltonian are sizeable, and the corresponding LEC can be constrained through the LQCD-FVEFT$_{\pislash}$ matching procedure. For the partial-wave mixing term, the analysis yields only an upper bound on its magnitude. 

This works demonstrates that FVEFT$_{\pislash}$ is a powerful alternative to the L\"uscher method~\cite{Luscher:1986pf} for analysis of systems of nucleons. There are two main advantages. On the one hand, the same scheme for the LECs of the EFT can be directly used in finite volume and in infinite volume---the later can allow for EFT calculations of larger nuclei than can be studied in LQCD at the present time. Second, FVEFT$_{\pislash}$ can be applied straightforwardly to systems of three and more nucleons, provided that LQCD spectra are available. By contrast, L\"uscher-like methods have only been derived for up to three particles~\cite{Hansen:2014eka,Hansen:2015zga,Hammer:2017uqm,Hammer:2017kms,Mai:2017bge}---see Refs.~\cite{Hansen:2019nir,Rusetsky:2019gyk,Romero-Lopez:2021zdo,Romero-Lopez:2022usb} for a review.

This application of FVEFT$_{\pislash}$ can be extended in several ways. For instance, higher orders in pionless EFT can be included. Moreover, other EFTs can be used, e.g., an EFT for baryons with strangeness, or chiral EFTs. As LQCD results become more precise and sophisticated, it will be important to investigate these extensions.

Ultimately, there is great potential for FVEFT analyses to extend the reach of LQCD calculations of light nuclei to constraints of the binding energies and matrix elements of larger nuclei than those that can be studied directly. Examples of important applications include scalar matrix elements and the isotensor axial polarisability~\cite{Shanahan:2017bgi}, which are relevant for dark matter and neutrino experiments~\cite{Davoudi:2020ngi,Cirigliano:2020yhp}. For this, direct constraints of LECs from LQCD at physical quark masses are needed, including those of three- and even four-body systems. This will require progress in both LQCD calculations, as well as in EFTs.

\acknowledgements

The authors thank Di Luo and Xiangkai Sun for useful discussions regarding this work. Special thanks is also due to the NPLQCD collaboration for providing the covariance matrix of the LQCD energies used in this work.

This work has been supported in part by the National Science Foundation under Cooperative Agreement PHY-2019786 (The NSF AI Institute for Artificial Intelligence and Fundamental Interactions, http://iaifi.org/) and by the USDOE, Office of Science, Office of Nuclear Physics, under grant Contract Numbers DE-SC0011090 and DE-SC0021006 and by the SciDAC5 award DE-SC0023116. FRL acknowledges financial support by the Mauricio and Carlota Botton Fellowship.

\onecolumngrid

\appendix

\section{Expressions for the matrix elements}
\label{app:matrix}

Here, expressions for the matrix elements of the different terms in the Hamiltonian in \Cref{eq:Hij} are provided (see \Cref{subsec:matels}). For $\left[\mathbb{N}\right]_{i j}$, $\left[\mathbb{K}^{(2)}\right]_{i j}$ and $\left[\mathbb{V}_2^\text{LO}\right]_{i j}$, see Appendix C of Ref.~\cite{Detmold:2021oro}. Note that the matrix elements of the potential and kinetic term are not labelled with superindices in Ref.~\cite{Detmold:2021oro}, but are the same quantities as in this work. Some results are reproduced here for completeness.

In the evaluation of matrix elements, fully symmetric wave functions are used---see \Cref{eq:Nsymm}. To simplify notation, the sum over permutations is omitted here. For example, in the case of the normalization,
\begin{equation}
    \left[\mathbb{N}\right]_{i j} = \sum_{\mathcal{P}\mathcal{P}'} \left[  \mathbb{N}_{\mathcal{P}\mathcal{P}'}\right]_{i j} \equiv \sum_{\mathcal{P}\mathcal{P}'} \big[  \widetilde{ \mathbb{N}}\big]_{i j},
    \label{eq:sumP}
\end{equation}
that is, $\widetilde{ \mathbb{N}}$ is short-hand notation for the permuted object.

\subsection{Normalization}

Using the factorization of the wave function in each coordinate, 
the expression can be split as:
\begin{equation}
    \big[  \widetilde{\mathbb{N}}\big]_{i j} = \prod_{\alpha=x,y,z} \big[  \widetilde{\mathbb{N}}^{(\alpha)}\big]_{i j}.
\end{equation}
Each term can then be evaluated as: 
\begin{equation}
  \big[  \widetilde{\mathbb{N}}^{(\alpha)}\big]_{i j}  =   \sqrt{\frac{(2 \pi)^{N_n}}{\operatorname{Det}\left[C_{i ; j }^{(\alpha)}\right]}} \sum_{\mathbf{b}^{(\alpha)}} \exp \left[-\frac{1}{2} \Omega_{i  ; j }^{(\alpha)}\right],
\end{equation}
where $N_n$ is the number of nucleons,
\begin{equation}
    \Omega_{i  ; j^{\prime}}^{(\alpha)}=\left(L \mathbf{b}^{(\alpha)}\right) \cdot A_{i }^{(\alpha)} \cdot\left(L \mathbf{b}^{(\alpha)}\right)+\left(L \mathbf{b}^{(\alpha)}+\mathbf{d}_{i }^{(\alpha)}\right) \cdot B_{i }^{(\alpha)} \cdot\left(L \mathbf{b}^{(\alpha)}+\mathbf{d}_{i }^{(\alpha)}\right)+\mathbf{d}_{j }^{(\alpha)} \cdot B_{j }^{(\alpha)} \cdot \mathbf{d}_{j }^{(\alpha)}-\mathbf{\Xi}_{i;j}^{(\alpha)} \cdot\left[C_{i ; j }^{(\alpha)}\right]^{-1} \cdot  \mathbf{\Xi}_{i;j}^{(\alpha)},
\end{equation}
and
\begin{align}
\mathbf{\Xi}_{i ; j }^{(\alpha)}&=L A_{i }^{(\alpha)} \cdot \mathbf{b}^{(\alpha)}+B_{i }^{(\alpha)} \cdot\left(L \mathbf{b}^{(\alpha)}+\mathbf{d}_{i }^{(\alpha)}\right)+B_{j } \cdot \mathbf{d}_{j }^{(\alpha)}, \\
    C_{i; j }^{(\alpha)}&=  C_i^{(\alpha)} + C_j^{(\alpha)}, \\
    C_i^{(\alpha)} &=  A_{i }^{(\alpha)}+B_{i }^{(\alpha)}.
\end{align}

\subsection{Nonrelativistic Kinetic term}

The evaluation of the matrix elements of the kinetic term can be split in different summands corresponding to the derivative operator for all particles in each direction:
\begin{equation}
    \left[\widetilde{\mathbb{K}}^{(2)}\right]_{i j} =  \frac{1}{2M_N} \sum_{\alpha=x,y,z}\left[\widetilde{\mathbb{K}}^{(2,\alpha)}\right]_{i j} \prod_{\beta \neq \alpha}  \big[  \widetilde{\mathbb{N}}^{(\beta)}\big]_{i j}.
\end{equation}
In the previous equation, the factorization of spatial parts has also been used. The remaining piece is
\begin{equation}
    \left[\widetilde{\mathbb{K}}^{(2,\alpha)}\right]_{i j} =  \sqrt{\frac{(2 \pi)^{N_n}}{\operatorname{Det}\left[C_{i ; j}^{(\alpha)}\right]}} \sum_{\mathbf{b}^{(\alpha)}}\Theta_{i ; j}^{(\alpha)}(\mathbb{1})\exp \left[-\frac{1}{2} \Omega_{i ; j }^{(\alpha)}\right],
\end{equation}
where $\mathbb{1}$ is an identity matrix and
\begin{align}
\begin{split}
   \Theta_{i ; j}^{(\alpha)}(\mathcal M)=\operatorname{Tr}\left[\left(A_{i }^{(\alpha)}+B_{i }^{(\alpha)}\right) \cdot\left[C_{i  ; j }^{(\alpha)}\right]^{-1} \cdot\left(A_{j }^{(\alpha)}+B_{j }^{(\alpha)}\right)\cdot\mathcal M\right] - \left(\mathcal Y_{ij}^{(\alpha)}\right)^T  \cdot \mathcal{M}  \cdot \mathcal Y^{(\alpha)}_{ij},
\end{split}
\end{align}
where $\mathcal M$ is a generic matrix, and
\begin{align}
\begin{split}
    \mathcal Y_{ij}^{(\alpha)}=&\left(A_{j }^{(\alpha)}+B_{j }^{(\alpha)}\right) \cdot\left[C_{i  ; j }^{(\alpha)}\right]^{-1} \cdot\left(B_{i }^{(\alpha)} \cdot\left(L \mathbf{b}^{(\alpha)}+\mathbf{d}_{i }^{(\alpha)}\right)+A_{i }^{(\alpha)} \cdot\left(L \mathbf{b}^{(\alpha)}\right)\right) \\
    &-\left(A_{i }^{(\alpha)}+B_{i }^{(\alpha)}\right) \cdot\left[C_{i ; j }^{(\alpha)}\right]^{-1} \cdot B_{j }^{(\alpha)} \cdot \mathbf{d}_{j}^{(\alpha)}.
\end{split}
\end{align}

\subsection{Total momentum}

The evaluation of the total momentum can be split in a similar way to the kinetic term:
\begin{equation}
    \left[\widetilde{\mathbb{P}}^{2}\right]_{i j} = \sum_{\alpha=x,y,z}\left[\widetilde{\mathbb{P}}^{2,(\alpha)}\right]_{i j} \prod_{\beta \neq \alpha}  \big[  \widetilde{\mathbb{N}}^{(\beta)}\big]_{i j},
\end{equation}
and in fact, the expression is very similar:
\begin{equation}
    \left[\widetilde{\mathbb{P}}^{(2,\alpha)}\right]_{i j} =  \sqrt{\frac{(2 \pi)^{N_n}}{\operatorname{Det}\left[C_{i ; j}^{(\alpha)}\right]}} \sum_{\mathbf{b}^{(\alpha)}}\Theta_{i ; j}^{(\alpha)}(\mathcal M_P)\exp \left[-\frac{1}{2} \Omega_{i ; j }^{(\alpha)}\right],
\end{equation}
where $\mathcal{M}_P$ is a matrix with all elements being unity, $\left[\mathcal{M}_P\right]_{nm} = 1$.

\subsection{Relativistic correction}

The relativistic correction matrix elements can be split in several terms:
\begin{equation}
    \left[\widetilde{\mathbb{K}}^{(4)}\right]_{i j} = \frac{-1}{8 M_N^3} \sum_{\alpha=x,y,z}\left[\widetilde{\mathbb{K}}^{(4,4,\alpha)}\right]_{i j} \prod_{\beta \neq \alpha}  \big[  \widetilde{\mathbb{N}}^{(\beta)}\big]_{i j}  - \frac{1}{4 M_N^3}  \sum_{n=1}^{N_n}\sum_{\alpha > \beta, \gamma\neq\alpha,\beta}   \left[\widetilde{\mathbb{K}}_n^{(4,2,\alpha)}\right]_{i j}    \left[\widetilde{\mathbb{K}}_n^{(4,2,\beta)}\right]_{i j}  \big[  \widetilde{\mathbb{N}}^{(\gamma)}\big]_{i j},
\end{equation}
where $\left[\widetilde{\mathbb{K}}_n^{(4,2,\beta)}\right]_{i j}$ is related to the nonrelativistic kinetic term for a single particle:
\begin{equation}
    \left[\widetilde{\mathbb{K}}_n^{(4,2,\alpha)}\right]_{i j} =  \sqrt{\frac{(2 \pi)^{N_n}}{\operatorname{Det}\left[C_{i ; j}^{(\alpha)}\right]}} \sum_{\mathbf{b}^{(\alpha)}}\Theta_{i ; j}^{(\alpha)}(\mathcal{M}_{K,n})\exp \left[-\frac{1}{2} \Omega_{i ; j }^{(\alpha)}\right],
\end{equation}
with $\left[ \mathcal{M}_{K,n} \right]_{l m} = \delta_{ln}\delta_{mn}$.  Finally, the last piece:
\begin{equation}
    \left[\widetilde{\mathbb{K}}^{(4,4,\alpha)}_n\right]_{i j} = 
    \sqrt{\frac{(2 \pi)^{N_n}}{\operatorname{Det}\left[C_{i ; j}^{(\alpha)}\right]}} \sum_{\mathbf{b}^{(\alpha)}}\left( \sum_{w=1}^6  \mathcal K^{(4,n,\alpha)}_{w,i;j}  \right) \exp \left[-\frac{1}{2} \Omega_{i ; j }^{(\alpha)}\right],
\end{equation}
where 
\begin{align}
\begin{split}
 \mathcal K^{(4,n,\alpha)}_{1,i;j} &= \sum_{klmo} \left[C_i^{(\alpha)}\right]_{nk} \left[C_i^{(\alpha)}\right]_{nl} \left[C_j^{(\alpha)}\right]_{nm} \left[C_j^{(\alpha)}\right]_{no} \left( \left[e_1\right]_{klmo} + \left[e_2\right]_{klmo} + \left[e_3\right]_{klmo} \right), \\
\mathcal K^{(4,n,\alpha)}_{2,i;j} &= \left(
\left( B_i^{(\alpha)} \mathbf{d}_i^{(\alpha)} + C_i^{(\alpha)} L \mathbf{b}^{(\alpha)} \right)_n^2
- \left[C^{(\alpha)}_i\right]_{nn}  \right) \left( \left(B_j^{(\alpha)} \mathbf{d}_j^{(\alpha)}\right)_n^2 -  \left[C^{(\alpha)}_j\right]_{nn} \right), \\ 
 \mathcal K^{(4,n,\alpha)}_{3,i;j}
&=  \sum_{kl}\left[\left(B_i^{(\alpha)}\mathbf{d}_i^{(\alpha)}+C_i^{(\alpha)} L \mathbf{b}^{(\alpha)} \right)_n^2 - \left[C^{(\alpha)}_i\right]_{nn} \right] (C^{(\alpha)}_j)_{nk} (C^{(\alpha)}_j)_{nl} \left( ( C\mathbf{\Xi})_k  (C\mathbf{\Xi})_l + \left[C_{i  ; j }^{(\alpha)\,-1}\right]_{kl} \right) \\
& \hspace{1cm} + \left[\left(B_j^{(\alpha)}\mathbf{d}_j^{(\alpha)} \right)_n^2 - \left[C^{(\alpha)}_j\right]_{nn} \right] (C^{(\alpha)}_i)_{nk} (C^{(\alpha)}_i)_{nl} \left( ( C\mathbf{\Xi})_k  (C\mathbf{\Xi})_l + \left[C_{i  ; j }^{(\alpha)\,-1}\right]_{kl} \right) , \\
 \mathcal K^{(4,n,\alpha)}_{4,i;j} &=-2 \sum_{klm} (C^{(\alpha)}_i)_{nk}(C^{(\alpha)}_i)_{nl} (C^{(\alpha)}_j)_{nm} \left(B_j^{(\alpha)}\mathbf{d}_j^{(\alpha)} \right)_n    \mathcal D_{klm}^{(\alpha),i;j} \\
                &\ \ \ - 2 \sum_{klm}
                (C^{(\alpha)}_j)_{nk}(C^{(\alpha)}_j)_{nl} (C^{(\alpha)}_i)_{nm}                 \left(B_i^{(\alpha)}\mathbf{d}_i^{(\alpha)}+C_i^{(\alpha)} L \mathbf{b}^{(\alpha)} \right)_n  
                  \mathcal D_{klm}^{(\alpha),i;j}, \\
 \mathcal K^{(4,n,\alpha)}_{5,i;j} &=  \sum_{k}- 2\left[\left(B_j^{(\alpha)}\mathbf{d}_j^{(\alpha)} \right)^2_n  - \left[C^{(\alpha)}_j\right]_{nn}  \right] \left(B_i^{(\alpha)}\mathbf{d}_i^{(\alpha)}+C_i^{(\alpha)} L \mathbf{b}^{(\alpha)} \right)_n  \left[C^{(\alpha)}_i\right]_{nk}  (C\mathbf{\Xi})_k \\
        & \hspace{0.9cm} - 2\left(B_j^{(\alpha)}\mathbf{d}_j^{(\alpha)} \right)_n \left[\left(B_i^{(\alpha)}\mathbf{d}_i^{(\alpha)}+C_i^{(\alpha)} L \mathbf{b}^{(\alpha)} \right)_n^2 - \left[C^{(\alpha)}_i\right]_{nn} \right] \left[C^{(\alpha)}_j\right]_{nk}  (C\mathbf{\Xi})_k,\\
  \mathcal K^{(4,n,\alpha)}_{6,i;j} &=  \sum_{kl} 4\left(B_j^{(\alpha)}\mathbf{d}_j^{(\alpha)} \right)_n \left(B_i^{(\alpha)}\mathbf{d}_i^{(\alpha)}+C_i^{(\alpha)} L \mathbf{b}^{(\alpha)} \right)_n (C^{(\alpha)}_i)_{nk} (C^{(\alpha)}_j)_{nl} \left( ( C\mathbf{\Xi})_k  (C\mathbf{\Xi})_l + \left[C_{i  ; j }^{(\alpha)\,-1}\right]_{kl} \right), \\
\end{split}
\end{align}
and shorthand notation is used such that 
\begin{equation}
 (C\mathbf{\Xi})_k  \equiv  \left( \left[C_{i  ; j }^{(\alpha)\,-1}\right] \mathbf{\Xi}_{i ; j }^{(\alpha)} \right)_k,
\end{equation}
as well as 
\begin{align}
\begin{split}
&\left[e_1\right]_{klmo} = \left[C_{i  ; j }^{(\alpha)\,-1}\right]_{kl} \left[C_{i  ; j }^{(\alpha)\,-1}\right]_{mo} + (l \leftrightarrow m) + (l \leftrightarrow o),  \\ 
 &\left[e_2\right]_{klmo} =  
  (C\mathbf{\Xi})_k
 (C\mathbf{\Xi})_l
 \left[C_{i  ; j }^{(\alpha)\,-1}\right]_{mo}
 + (k \leftrightarrow m) + (k \leftrightarrow o) 
 + (l \leftrightarrow m) + (l \leftrightarrow o) 
 + (l,k \leftrightarrow m,o ),
 \\
 &\left[e_3\right]_{klmo} =   (C\mathbf{\Xi})_k  (C\mathbf{\Xi})_l  (C\mathbf{\Xi})_m  (C\mathbf{\Xi})_o,
\end{split}
\end{align}
and 
\begin{equation}
    \mathcal D_{klm}^{(\alpha),i;j} = \left[C_{i  ; j }^{(\alpha)\,-1}\right]_{kl} (C\mathbf{\Xi})_m + \left[C_{i  ; j }^{(\alpha)\,-1}\right]_{km} (C\mathbf{\Xi})_l + \left[C_{i  ; j }^{(\alpha)\,-1}\right]_{ml} (C\mathbf{\Xi})_k +  (C\mathbf{\Xi})_k (C\mathbf{\Xi})_l (C\mathbf{\Xi})_m.
\end{equation}

\subsection{Leading-order potential}

The matrix element for the LO potential can be split as:
\begin{equation}
    \left[\widetilde{\mathbb{V}}_{2}\right]_{i j} = \prod_{\alpha}  \big[  \widetilde{\mathbb{V}}_2^{(\alpha)}\big]_{i j},
\end{equation}
where
\begin{align}
\begin{split}
    \big[  \widetilde{\mathbb{V}}_2^{(\alpha)}\big]_{i j} &= \sqrt{\frac{(2 \pi)^{N_n}}{\operatorname{Det}\left[C_{i  ; j }^{(\alpha)}\right]}} \sqrt{\frac{\widetilde{C}_{i  ; j }^{(\alpha)}}{\widetilde{C}_{i  ; j }^{(\alpha)}+2 \rho} }
\sum_{\mathbf{b}^{(\alpha)}} \exp \left[-\frac{1}{2} \Omega_{i  ; j }^{(\alpha)}\right] \\
&\times \sum_{q^{(\alpha)}} \exp \left[-\frac{\rho \,\widetilde{C}_{i ; j }^{(\alpha)}}{\widetilde{C}_{i  ; j }^{(\alpha)}+2 \rho}\left((C \Xi)_n-(C \Xi)_m-L q^{(\alpha)}\right)^2\right],
\end{split}
\end{align}
with $\rho={1}/(2 r_0^2)$ and
\begin{equation}
    \widetilde{C}_{i  ; j }^{(\alpha)}=\left(\left[C_{i ; j }^{(\alpha)\,-1}\right]_{n n}+\left[C_{i  ; j }^{(\alpha)\,-1}\right]_{m m}-\left[C_{i ; j}^{(\alpha)\,-1} \right]_{n m}-\left[C_{i  ; j}^{(\alpha)\,-1} \right]_{m n}\right)^{-1}.
\end{equation}
Here $n$ and $m$ correspond to the indices of the particles that are interacting, see \Cref{eq:hamKV}. For a two-nucleon system, $n=1$, $m=2$. The opposite $m=1$, $n=2$ combination is included in the sum over permutation as in \Cref{eq:sumP}.

\subsection{Next-to-leading-order potential}

The matrix element of the NLO potential can be computed as:
\begin{equation}
\left[\widetilde{\mathbb{V}}^\text{NLO}_{2}\right]_{i j} = \sum_\alpha  \big[  \widetilde{\mathbb{V}}_2^{\text{NLO},(\alpha)}\big]_{i j}
    \prod_{\beta\neq\alpha}  \big[  \widetilde{\mathbb{V}}_2^{(\beta)}\big]_{i j},
\end{equation}
where
\begin{align}
\begin{split}
    \big[  \widetilde{\mathbb{V}}_2^{\text{NLO},(\alpha)}\big]_{i j} &= \sqrt{\frac{(2 \pi)^{N_n}}{\operatorname{Det}\left[C_{i  ; j }^{(\alpha)}\right]}} \sqrt{\frac{\widetilde{C}_{i  ; j }^{(\alpha)}}{\widetilde{C}_{i  ; j }^{(\alpha)}+2 \rho} }
\sum_{\mathbf{b}^{(\alpha)}} \exp \left[-\frac{1}{2} \Omega_{i  ; j }^{(\alpha)}\right] 
\\&\times \sum_{q^{(\alpha)}} \left[ \text{Tr }\left( \left(C^{(\alpha)}_i + 2\rho \mathcal M_V^{(mn)}\right) \cdot \left(C_{i ; j }^{(\alpha)} + 2\rho \mathcal M_V^{(mn)}\right)^{-1} \cdot C^{(\alpha)}_j\cdot \mathcal M^{(nm)}_V\right)  - (\mathcal{Y}^{(\alpha),Q}_{ij})^T\cdot \mathcal M^{(nm)}_V \cdot\mathcal{Y}^{(\alpha),Q}_{ij}  \right] \\ 
&\times \exp \left[-\frac{\rho \,\widetilde{C}_{i ; j }^{(\alpha)}}{\widetilde{C}_{i  ; j }^{(\alpha)}+2 \rho}\left((C \Xi)_n-(C \Xi)_m-L q^{(\alpha)}\right)^2\right],
\end{split}
\end{align}
where $n=1$ and $m=2$ are used, and
\begin{align}
\begin{split}
    \mathcal Y^{(\alpha),Q}_{ij}=&\left(A_{j }^{(\alpha)}+B_{j }^{(\alpha)}\right) \cdot\left[C_{i  ; j }^{(\alpha)}+ 2\rho \mathcal M_V^{(mn)}\right]^{-1} \cdot\left(B_{i }^{(\alpha)} \cdot\left(L \mathbf{b}^{(\alpha)}+\mathbf{d}_{i }^{(\alpha)}\right)+A_{i }^{(\alpha)} \cdot\left(L \mathbf{b}^{(\alpha)}\right)\right) \\
    &-\left(A_{i }^{(\alpha)}+B_{i }^{(\alpha)}+ 2\rho \mathcal M_V^{(mn)}\right) \cdot\left[C_{i ; j }^{(\alpha)}+ 2\rho \mathcal M_V^{(mn)}\right]^{-1} \cdot B_{j }^{(\alpha)} \cdot \mathbf{d}_{j}^{(\alpha)}+2\rho \, L q^{(\alpha)}  \mathbf w^{(nm)},
\end{split}
\end{align}
with $\mathbf w^{(nm)}_k= \delta_{kn} -\delta_{km}$, and $(\mathcal M^{(nm)}_V)_{kl} = \left( \delta_{kn}\delta_{ln} + \delta_{km}\delta_{lm} - \delta_{kn}\delta_{lm} - \delta_{km}\delta_{ln}  \right) $.

Note that in \Cref{eq:V2NLOpot}, two terms are needed where the derivatives act on different sides. The above expressions correspond to the derivatives acting on the right side, and the other case is obtained by switching $i\leftrightarrow j$. Note that because of the sum over permutations in \Cref{eq:sumP}, it is not necessary to switch $n \leftrightarrow m$.  

\subsection{$s$-$d$-wave mixing term}

The $s$-$d$-wave mixing term is computed as:
\begin{equation}
\left[\widetilde{\mathbb{V}}^{sd}_2\right]_{i j} = \sum_{\alpha\beta\gamma\delta} \mathcal{T}^{\gamma\delta}_{\alpha\beta} 
\left[{\widetilde{\mathbb{V}}}^{(\alpha\beta),sd}_{2}\right]_{i j}.
\end{equation}
The component of the spatial part with identical indices is given by:
\begin{equation}
\left[{\widetilde{\mathbb{V}}}^{(\alpha\alpha),sd}_{2}\right]_{i j} = \big[  \widetilde{\mathbb{V}}_2^{\text{NLO},(\alpha)}\big]_{i j}
    \prod_{\beta\neq\alpha}  \big[  \widetilde{\mathbb{V}}_2^{(\beta)}\big]_{i j}.
\end{equation}
The off-diagonal part ($\alpha\neq\beta$):
\begin{equation}
\left[{\widetilde{\mathbb{V}}}^{(\alpha\beta),sd}_{2}\right]_{i j} =\sum_{k \in \{m,n\}} \sum_{l \in \{m,n\}}  \left( (\mathcal M^{(nm)}_V)_{kl} \big[  \widetilde{\mathbb{V}}_2^{sd,(k,\alpha)}\big]_{i j}
\big[  \widetilde{\mathbb{V}}_2^{sd,(l,\beta)}\big]_{i j} \right) \prod_{\gamma\neq \alpha, \beta}
    \big[  \widetilde{\mathbb{V}}_2^{(\gamma)}\big]_{i j},
\end{equation}
 and
\begin{align}
\begin{split}
    \big[  \widetilde{\mathbb{V}}_2^{sd,(k,\alpha)}\big]_{i j} &= \sqrt{\frac{(2 \pi)^{N_n}}{\operatorname{Det}\left[C_{i  ; j }^{(\alpha)}\right]}} \sqrt{\frac{\widetilde{C}_{i  ; j }^{(\alpha)}}{\widetilde{C}_{i  ; j \mathcal{P}^{\prime}}^{(\alpha)}+2 \rho} }
\sum_{\mathbf{b}^{(\alpha)}} \exp \left[-\frac{1}{2} \Omega_{i  ; j }^{(\alpha)}\right] 
\\&\times \left(\mathcal{Y}^{(\alpha),Q}_{ij}\right)_k
 \exp \left[-\frac{\rho \,\widetilde{C}_{i ; j }^{(\alpha)}}{\widetilde{C}_{i  ; j }^{(\alpha)}+2 \rho}\left((C \Xi)_n-(C \Xi)_m-L q^{(\alpha)}\right)^2\right].
\end{split}
\end{align}

Note that in \Cref{eq:sdpotential}, two terms are needed where the derivatives act on different sides. The above expressions correspond to the derivatives acting on the right side, and the other case is obtained by switching $i\leftrightarrow j$. Note that because of the sum over permutations in \Cref{eq:sumP}, it is not necessary to switch $n \leftrightarrow m$.

\bibliography{reference}
\end{document}